\documentclass[aps,pre,superscriptaddress]{revtex4}
\usepackage[dvips]{graphics}
\usepackage{graphicx}
\usepackage{amsfonts}
\usepackage{amssymb}
\usepackage{amsmath}
\usepackage{subfigure}
\usepackage[abs]{overpic}

\begin{document}

\title{Linear and Nonlinear PT-symmetric Oligomers: A Dynamical Systems
Analysis}

\author{M. Duanmu}
\affiliation{Department of Mathematics and Statistics, University of Massachusetts,
Amherst MA 01003-4515, USA}

\author{K. Li}
\affiliation{Department of Mathematics and Statistics, University of Massachusetts,
Amherst MA 01003-4515, USA}

\author{R.L. Horne}
\affiliation{Department of Mathematics, Morehouse College, Atlanta, GA 30314}
\email{rhorne@morehouse.edu}

\author{P.G.\ Kevrekidis }
\affiliation{Department of Mathematics and Statistics, University of Massachusetts,
Amherst MA 01003-4515, USA}
\email{kevrekid@math.um}

\author{N. Whitaker}
\affiliation{Department of Mathematics and Statistics, University of Massachusetts,
Amherst MA 01003-4515, USA}
\email{whitaker@math.umass.edu}

\begin{abstract}
In the present work we focus on the cases of two-site (dimer) and three-site (trimer) configurations, i.e. oligomers, respecting the parity-time (PT) symmetry,
i.e., with a spatially odd gain-loss profile. We examine different types of
solutions of such configurations with linear and nonlinear gain/loss profiles.
Solutions beyond the linear PT-symmetry critical point as well as solutions
with asymmetric linearization eigenvalues are found in both the
nonlinear dimer and trimer. The latter feature is absent in linear
PT-symmetric trimers, while both of them are absent in linear
PT-symmetric dimers. Furthermore, nonlinear gain/loss terms enable the
existence of both symmetric and asymmetric solution profiles (and of
bifurcations between them), while only symmetric solutions are present in the
linear PT-symmetric dimers and trimers.
The linear stability analysis around the obtained solutions is discussed and
their dynamical evolution is explored by means of direct numerical simulations.
Finally, a brief discussion is also given of recent progress in the
context of PT-symmetric quadrimers.
\end{abstract}

\maketitle

\section{Introduction}

In the late 1990s, a radical yet well physically motivated proposal
emerged in the context of the study of fundamentals of quantum mechanics.
This was the suggestion of Bender and co-workers~\cite{bend} that
Hamiltonians that respect two principal physical symmetries of the
dynamics, namely Parity (P) and Time-reversal (T) could enable
the identification of real eigenvalues, which is a property which
is highly desirable for operators associated with measurable
quantities, even if the associated Hamiltonians are not Hermitian.
A caveat in that regard, however, was that as the gain/loss parameter
introduced in these non-Hermitian Hamiltonians was varied, a
transition was quantified and termed the PT-symmetry breaking
transition, beyond which the eigenvalues of the relevant
operator were no longer purely real. It is interesting to note
here that a prototypical playground where such ideas can be
explored is that of standard Schr{\"o}dinger
Hamiltonians of the form $H=-(1/2) \Delta + V(x)$, for which
in the case of a complex potential $V$, it is straightforward to see
that the above constraints of PT symmetry amount to the potential satisfying
the condition  $V(x)=V^{\star}(-x)$.

While the initial proposal of the above possibility was one of
mathematical origin (rather than one inspired by a specific physical
application), in the past few years, a significant vein of potential
applications of such Hamiltonians has been initiated, predominantly
so in the field of nonlinear optics. There, the work of
Christodoulides and co-workers~\cite{christo1} gave rise to the
realization that the synthetic systems that can be engineered
therein enable a potential balance of gain (through suitable
amplifiers) and the abundantly present losses in order to
produce experimental realizations of PT-symmetric systems.
An additional feature present in such settings which
made both their theoretical and experimental investigation
even more interesting was the presence of nonlinearity
which, in turn, rendered worthwhile the exploration of the
dynamics of nonlinear waves (such as bright or gap solitons~\cite{christo1}
and more recently of dark solitons and vortices~\cite{usrecent}).
The above optical settings were in fact the ones that
enabled the first experimental realizations of PT-symmetry.
This was done in the context of waveguide couplers (i.e., either
two waveguides with and without loss~\cite{salamo} -- the so-called
passive PT-- or in the more ``standard'' case of one waveguide
with gain and one with loss~\cite{kip}). More recently,
electronic analogs of such systems have been engineered in
the work of~\cite{tsampikos_recent}. In parallel to these pioneering
steps in the realm of experiments, numerous theoretical
investigations have arisen both in the context of linear
PT-symmetric
potentials~\cite{kot1,sukh1,kot2,grae1,grae2,kot3,pgk,dmitriev1,dmitriev2,R30add1,R30add2,R30add3,R30add4,R30add5,R34,R44,R46}
and even in the case of the so-called nonlinear PT-symmetric potentials
(whereby a PT-symmetric type of gain/loss pattern appears in the nonlinear
term)~\cite{miron,konorecent,konorecent2}.

In the present work, we revisit this theme of both linear and nonlinear
PT-symmetric potentials. We do so in the special context of the recently
proposed PT-symmetric ``oligomers'', namely few site configurations, which
are not necessarily couplers (i.e., dimers) but rather can also be
trimers, quadrimers etc. For the sake of illustration, we restrict
ourselves to dimers and trimers herein but briefly also touch upon
recent works in the context of quadrimers. In what follows below,
we study the case of nonlinear-PT-symmetric dimers~\cite{miron} and trimers
and subsequently restrict considerations to the case of their
linear-PT-symmetric analogs~\cite{pgk}. One of the fundamental side-effects
of the fact that the nonlinearity does {\it not} commute with the PT
operator is the existence of nonlinear solutions that persist past the
linear PT-phase-transition threshold.
Furthermore, one of the principal consequences of the presence
of gain/loss terms in the nonlinearity is the existence of both symmetric
and asymmetric (in their amplitude) stationary solutions, with the latter
possessing a non-symmetric linearization spectrum. Interesting bifurcation
phenomena (such as spontaneous symmetry breakings)
are, additionally, found to arise in this case. Our presentation
is structured as follows: we first examine the nonlinear-PT-symmetric
dimer in section II (existence of solutions in II.A, linear stability setup
in II.B and numerical results in II.C), while the corresponding analysis
for the trimer is done in section III. In section IV, we review the
corresponding case of linear-PT-symmetric oligomers, while in section V,
we summarize our findings and present our conclusions.

\section{Analysis of Stationary  Solutions for the Nonlinear-PT-Symmetric
Dimer Case}

We first consider the so-called PT-symmetric coupler or dimer, in which
a gain/loss pattern appears {\it both} in the linear and
nonlinear terms. The dynamical equations
have the form:
\begin{eqnarray} \label{basic_dimer_equations}
iu_{t} = -\epsilon v + (\rho_{r} - i\rho_{im})|u|^{2}u + i\gamma u \nonumber \\
    \nonumber  \\
iv_{t} = -\epsilon u + (\rho_{r} + i\rho_{im})|v|^{2}v - i\gamma v.
\end{eqnarray}
The
model contains the Kerr nonlinearity which is relevant to optical waveguides
and is effectively a generalization of the experimental framework
of~\cite{kip}, in that nonlinear (i.e., amplitude-dependent)
gain and loss processes are taken into account. We have used $u(t)$
and $v(t)$ to denote the two complex-valued variables for the dimer and
the evolution variable is $t$ (in optics, this is actually a spatial
variable standing for the propagation distance along the optical crystal).
Considering the prototypical stationary solutions of the system, we let
$u(t)$ and $v(t)$ have the forms:
\begin{eqnarray}
  u(t) = ae^{iEt}, ~~~~~~ v(t) = be^{iEt}
\end{eqnarray}
where $E$ is the propagation constant
while the complex numbers $a$ and $b$ denote the amplitudes of the dimer
sites. Plugging this
ansatz into Eq.~($\ref{basic_dimer_equations}$), one finds the complex
nonlinear algebraic equations:
\begin{eqnarray} \label{ab_dimer_linear_equations}
  Ea = \epsilon b - (\rho_{r} - i\rho_{im})|a|^{2}a - i\gamma a \nonumber \\
       \nonumber \\
 Eb = \epsilon a - (\rho_{r} + i\rho_{im})|b|^{2}b + i\gamma b.
\end{eqnarray}
We now use a polar decomposition of $a$ and $b$ of the form:
\begin{eqnarray} \label{ab_dimer_forms}
 a = Ae^{i\phi_{a}}, ~~~~~~ b = Be^{i\phi_{b}}
\end{eqnarray}
for real-valued $A$, $B$, $\phi_{a}$ and $\phi_{b}$. Plugging Eq.~($\ref{ab_dimer_forms}$) into
Eq.~($\ref{ab_dimer_linear_equations}$) and writing these equations in terms of their real and imaginary parts,
we find:
\begin{eqnarray} \label{ab_complex_dimer}
EA = \epsilon B \cos(\phi_{b} - \phi_{a}) - \rho_{r} A^{3} \nonumber \\
     \nonumber \\
EB = \epsilon A \cos(\phi_{b} - \phi_{a}) - \rho_{r} B^{3} \nonumber \\
    \nonumber \\
 - \epsilon A \sin(\phi_{b} - \phi_{a}) - \rho_{im} B^{3} + \gamma B = 0 \\
    \nonumber \\
 \epsilon B \sin(\phi_{b} - \phi_{a}) + \rho_{im} A^{3} - \gamma A = 0. \nonumber
\end{eqnarray}
The last two equations yield
\begin{eqnarray} \label{AB_dimer_cases}
    (A^{2} - B^{2})\left[ \rho_{im}(A^{2} + B^{2}) - \gamma \right] = 0.
\end{eqnarray}
We note that Eq.~($\ref{AB_dimer_cases}$) yields a simple algebraic
condition which connects the amplitude of the two dimer sites. This
allows us to distinguish several subcases of interest. We look for nontrivial
solutions $A$ and $B$ in each of
the subcases presented in the following section.

\subsection{Existence of Localized Modes for the Dimer Case}
Eq.~($\ref{AB_dimer_cases}$)
identifies the different scenarios for the values of $A$ and $B$.
We now examine the three cases that arise from this equation for our dimer dynamical system.
\begin{itemize}
   \item $\underline{\mbox{Case I}:~~A^{2} = B^{2} ~~\mbox{and}~~ A^{2} + B^{2} \neq \gamma/\rho_{im}}$:\\
   \\
     Recall the equations given in~($\ref{ab_complex_dimer}$):
     \begin{eqnarray}
         EA = \epsilon B \cos(\phi_{b} - \phi_{a}) - \rho_{r} A^{3}; ~~~~~~EB = \epsilon A \cos(\phi_{b} - \phi_{a}) - \rho_{r} B^{3}
     \end{eqnarray}
     and
     \begin{eqnarray}
         \epsilon B \sin(\phi_{b} - \phi_{a}) + \rho_{im}A^{3} - \gamma A = 0; ~~~~~~
          -\epsilon A \sin(\phi_{b} - \phi_{a}) - \rho_{im}B^{3} + \gamma B = 0.
     \end{eqnarray}
     Since $A = B$ (i.e., these are symmetric solutions) in this case,
the two equations in each set are equivalent. Thus, we have:
      \begin{eqnarray}
         \sin(\phi_{b} - \phi_{a}) = \frac{-(\rho_{im}A^{2} - \gamma)}{\epsilon}, ~~~~~~
         \cos(\phi_{b} - \phi_{a}) = \frac{\rho_{r}A^{2} + E}{\epsilon}.
      \end{eqnarray}
     We use the relation $\sin^{2}(\phi_{b} - \phi_{a}) + \cos^{2}(\phi_{b} - \phi_{a}) =1$ to determine the
     following quadratic equation for $A^{2}$:
       \begin{eqnarray}
          (\rho_{r}^{2} + \rho_{im}^{2})A^{4} + 2(E\rho_{r} - \gamma \rho_{im})A^{2} + \gamma^{2} + E^{2} - \epsilon^{2} = 0.
       \end{eqnarray}
The solution of the resulting bi-quadratic equation reads:
       \begin{eqnarray} \label{Asquared_A_equal_B_dimer_case}
          A^{2} = B^{2} = \frac{-(E\rho_{r} - \gamma \rho_{im}) \pm
           \sqrt{(E\rho_{r} - \gamma \rho_{im})^{2} - (\rho_{r}^{2} + \rho_{im}^{2})(\gamma^{2} + E^{2} - \epsilon^{2})}}{\rho_{r}^{2} + \rho_{im}^{2}},
       \end{eqnarray}
      with the restriction that


        \begin{eqnarray}\label{dimer1_restriction}
           (E\rho_{r} - \gamma \rho_{im})^{2} \ge (\rho_{r}^{2} + \rho_{im}^{2})(\gamma^{2} + E^{2} - \epsilon^{2}).
        \end{eqnarray}


  \item $\underline{ \mbox{Case II}:~~A^{2} +  B^{2} = \gamma/\rho_{im} ~~\mbox{and}~~ A^{2} \neq B^{2}}$: \\
       \\
        Under these conditions, one can get
        \begin{eqnarray}\label{dimer2}
        A^{2}&=&\frac{\gamma}{2\rho_{im}}\pm\sqrt{\frac{\gamma^2}{4\rho_{im}^{2}}-\frac{\epsilon^{2}}{\rho_{r}^{2}+\rho_{im}^{2}}} \\
        B^{2}&=&\dfrac{\gamma}{2\rho_{im}}\mp\sqrt{\frac{\gamma^2}{4\rho_{im}^{2}}-\frac{\epsilon^{2}}{\rho_{r}^{2}+\rho_{im}^{2}}} \\
        E &=& -\frac{\gamma\rho_r}{\rho_{im}}
\label{en_condition}
\\
        \cos(\phi_b-\phi_a) &=& -\frac{\epsilon\rho_r}{\sqrt{\rho_r^2 + \rho_{im}^2}},
        \end{eqnarray}
        with the restriction that
        \begin{eqnarray}
        \frac{\gamma^2}{4\rho_{im}^{2}}\ge\frac{\epsilon^{2}}{\rho_{r}^{2}+\rho_{im}^{2}}.
        \label{restrict_2}
        \end{eqnarray}
A fundamental difference of this case from case I is that
here $E$ is no longer a free parameter~\cite{miron}.
The solutions with the different amplitudes
will be called asymmetric in what follows.

  \item $\underline{ \mbox{Case III}:~~A^{2} +  B^{2} = \gamma/\rho_{im} ~~\mbox{and}~~ A^{2} = B^{2}}$:\\
        \\

        As a final ``mixed'' possibility, between the above
symmetric and
asymmetric cases, from
Eq.~(\ref{ab_complex_dimer}), it is straightforward to obtain
      \begin{eqnarray} \label{A_equal_B_dimer_case}
         A &=& B = \sqrt{\frac{\gamma}{2\rho_{im}}},
      \end{eqnarray}
      \begin{eqnarray} \label{cosine_sine_relation_dimer}
          \cos(\phi_{b} - \phi_{a}) &=& \frac{2\rho_{im}E + \gamma \rho_{r}}{2\epsilon \rho_{im}} \\
        \label{sine_relation}
          \sin(\phi_{b} - \phi_{a}) &=& \frac{\gamma}{2\epsilon},
       \end{eqnarray}
       with the restriction that
       \begin{eqnarray}
           \left(\frac{2\rho_{im}E + \gamma \rho_{r}}{2\epsilon \rho_{im}}\right)^{2} + \left(\frac{\gamma}{2\epsilon}\right)^{2} = 1.
\label{energy_eq}
       \end{eqnarray}
\end{itemize}
Once again this implies that
once other parameters (such as
$\gamma$, $\rho_{im}$, $\rho_{r}$ and $\epsilon$ are determined,
$E$ is not a free parameter but rather is obtained from Eq.~(\ref{energy_eq}).
These will be referred to as special symmetric solutions in the following.

It is particularly important to highlight that both solutions of Case II
(asymmetric) and ones of Case III (special symmetric) are present due to
competing effects of the linear and nonlinear gain loss profiles; notice
the opposite signs thereof in Eq.~(\ref{basic_dimer_equations}) and
the necessity of $\gamma \rho_{im} > 0$ for such solutions to exist.
In the case, where the linear and nonlinear gain/loss cooperate
(rather than compete) such solutions would obviously be absent
and the system would be inherently less wealthy in its potential
dynamics. This point was also discussed in~\cite{miron}.

 \subsection{Linear Stability Analysis for the Dimer Case}
%

We now go back to our original PT-symmetric dimer with
linear and nonlinear gain and
loss in Eq.~(\ref{basic_dimer_equations}) and examine the
linear stability of the
solutions to this equation. We begin by setting
  \begin{eqnarray} \label{uv_stability_solutions}
      u(t) = e^{iEt}[ a + pe^{\lambda t} + Pe^{\lambda^{*}t}], ~~~~~v(t) = e^{iEt}[ b + qe^{\lambda t} + Qe^{\lambda^{*}t}]
  \end{eqnarray}
  where $\lambda$ is a complex-valued eigenvalue parameter revealing the
growth (instability) or oscillation (stability) of all the modes of
linearization of the dimer system;
  ${*}$ denotes the complex conjugate and $p, P, q, Q$ are perturbations to
the solutions of interest. Plugging
  Eq.~($\ref{uv_stability_solutions}$) into Eq.~(\ref{basic_dimer_equations})
and taking only the linear terms
  in $p, P, q$ and $Q$, we find the following eigenvalue problem:
    \begin{eqnarray}
       {\bf A}{\bf X} = i\lambda {\bf X}
    \end{eqnarray}
    where ${\bf X} = (p, ~-P^{*}, ~q, ~-Q^{*})^{T}$ and ${\bf A}$ is written as:

   \begin{eqnarray}
          {\bf A} = \left( \begin{array}{cccc}
                                    a_{11} & -a^{2}(\rho_{r} - i\rho_{im}) &  -\epsilon & 0\\
                                    (a^{*})^{2}(\rho_{r} + i\rho_{im}) &a_{22}&0 & \epsilon\\
                                    -\epsilon & 0 & a_{33} & -b^{2}(\rho_{r} + i\rho_{im})\\
                                    0 &\epsilon& (b^{*})^{2}(\rho_{r} - i\rho_{im}) &  a_{44}
                            \end{array}\right)
                             \end{eqnarray}
    where
    \begin{eqnarray}
    a_{11} = E + 2|a|^{2}(\rho_{r} - i\rho_{im}) + i\gamma 
           \nonumber\\
             \nonumber \\
           a_{22} = -2|a|^{2}(\rho_{r} + i\rho_{im}) - E + i\gamma
           \nonumber\\
                \nonumber \\
           a_{33} = E - i\gamma + 2|b|^{2}(\rho_{r} + i\rho_{im})
            \nonumber\\
                 \nonumber \\
            ~~~a_{44} = -E - i\gamma - 2|b|^{2}(\rho_{r} - i\rho_{im}).
\end{eqnarray}
The use of the symmetric, asymmetric or mixed solutions of
the previous subsection into these matrix elements produces a $4 \times 4$
complex matrix whose eigenvalues will determine the spectral stability
of the corresponding nonlinear solution. The existence of eigenvalues with
positive real part $\lambda_r>0$ amounts to a dynamical instability of
the relevant solution, while in the case where all 4 eigenvalues have
$\lambda_r \leq 0$, the solution is linearly stable.

    \subsection{Numerical Results for the Dimer Case}
\begin{figure}[htp]
\scalebox{0.5}{\includegraphics{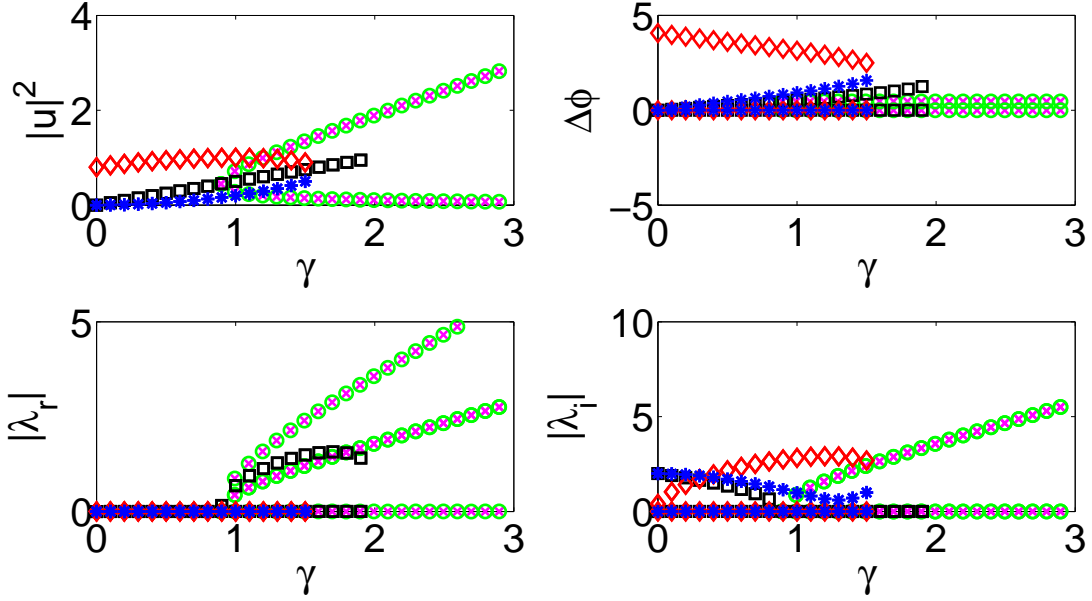}}
\caption{The solution profiles of the nonlinear PT-symmetric dimer
case with $\epsilon=1$, $\rho_r=-2$ and $\rho_{im}=1$.
The four panels here present the continuation of each branch
(the amplitudes
in the top left, the phases in the top right, and the real -bottom left-
and imaginary -bottom right- parts
of the linear
stability eigenvalues) starting from the conservative system at $\gamma=0$.
The five branches are denoted by curves of blue stars, red diamonds, black
squares, green circles and magenta crosses.
Blue stars: Case I with ``-'' in the amplitude;
Red diamonds: Case I with ``+'' in the amplitude;
Green circles: Case II with ``+'' in the amplitude (of $A$);
Magenta crosses: Case II with ``-'' in the amplitude (of $A$);
Black squares: Case III. Notice that
the eigenvalues of green circles and magenta crosses are opposite to each
other (see the relevant discussion in the text).
We always set $E=1$ in the case I branches, namely the blue stars and the red diamonds,
which terminate at the same point when
$\gamma=1.61$.
The black squares are subject to a destabilizing
supercritical pitchfork bifurcation at
$\gamma=0.895, E=1.789$ whereby the green circles and magenta crosses arise.
The black squares branch terminates at $\gamma=2$; the green circles and
magenta crosses exist for arbitrary values of the (linear) gain/loss past
the linear PT-symmetry breaking point.}
\label{fig4}
\end{figure}

\begin{figure}[htp]
\scalebox{0.38}{\includegraphics{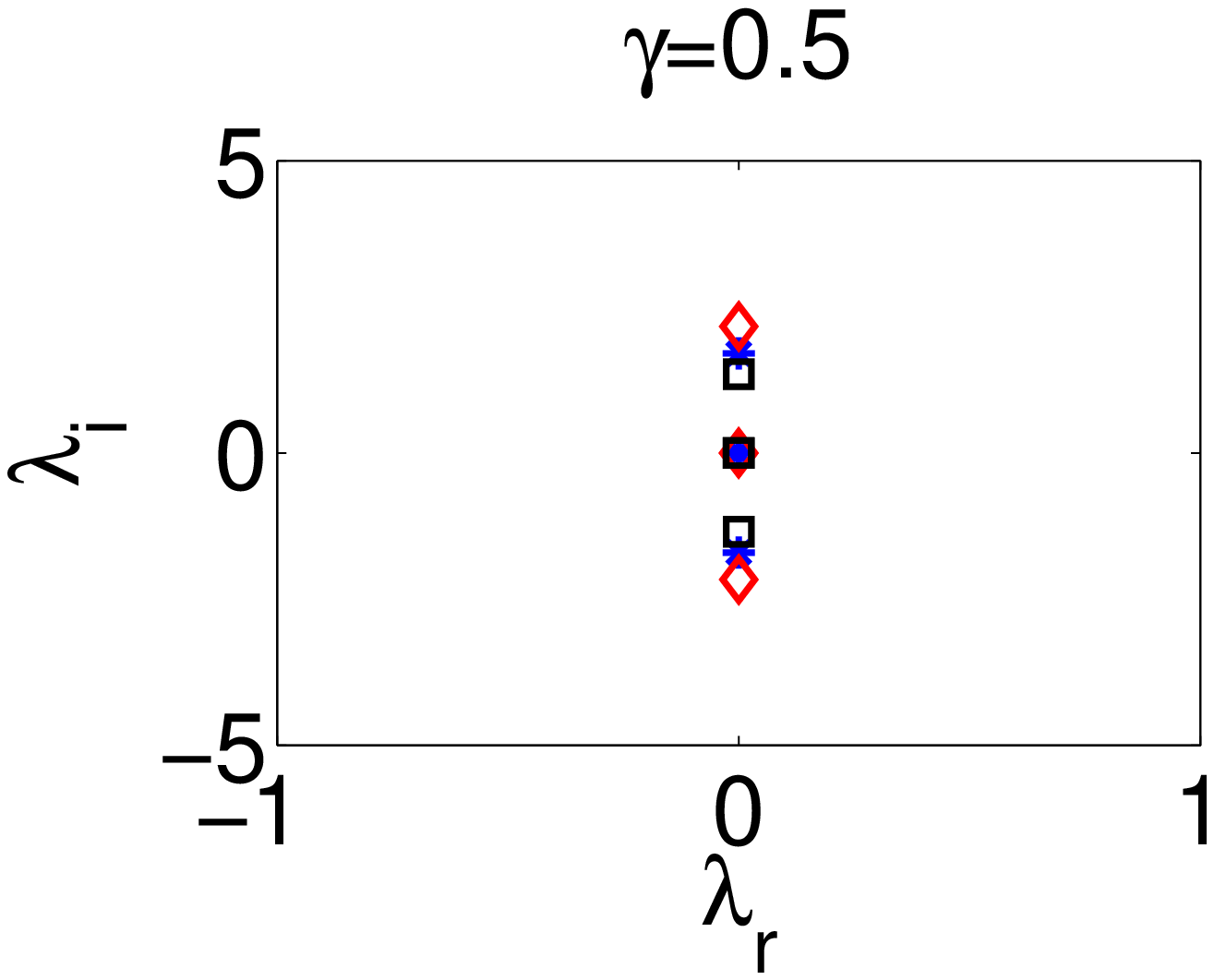}}
\scalebox{0.38}{\includegraphics{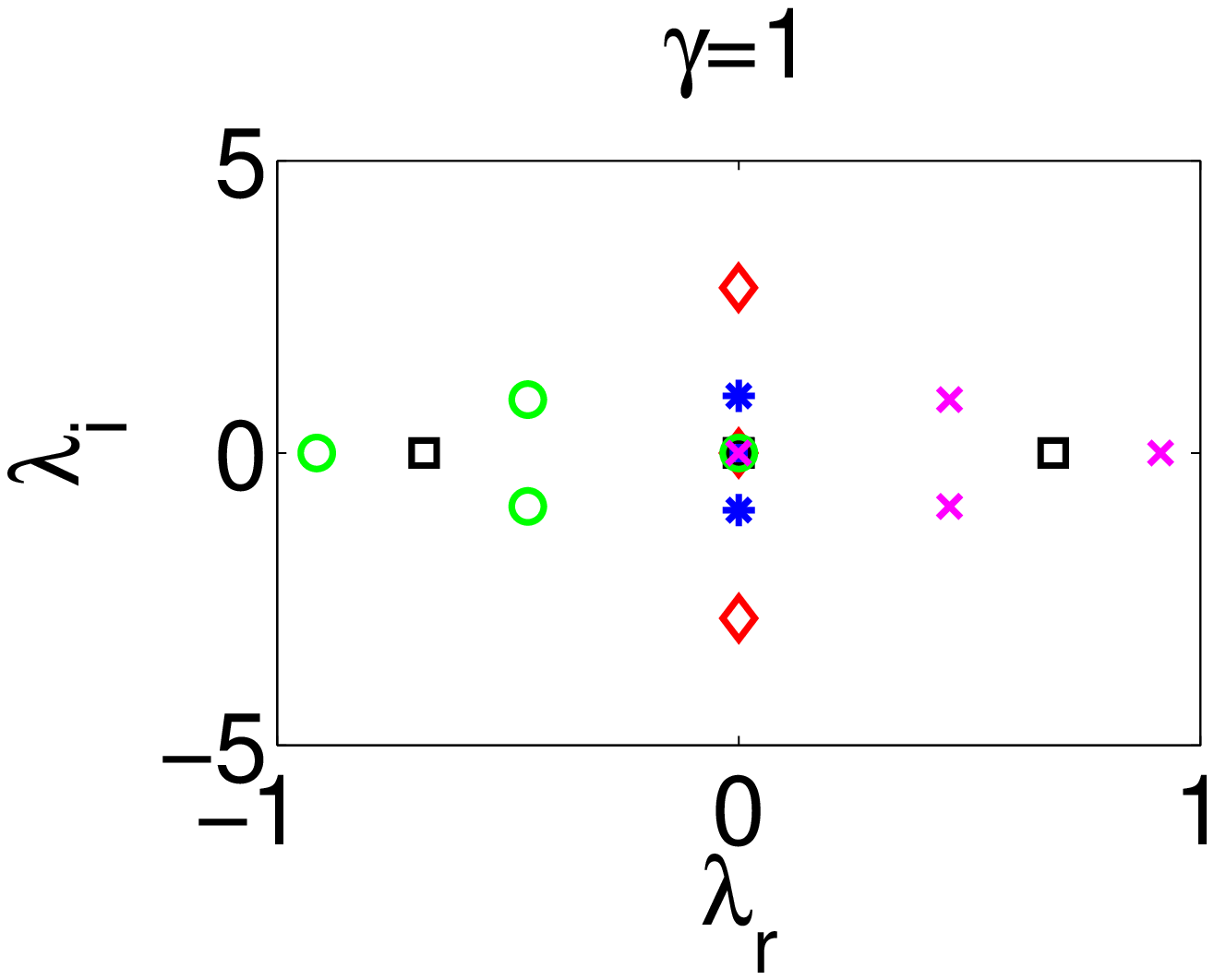}}
\scalebox{0.38}{\includegraphics{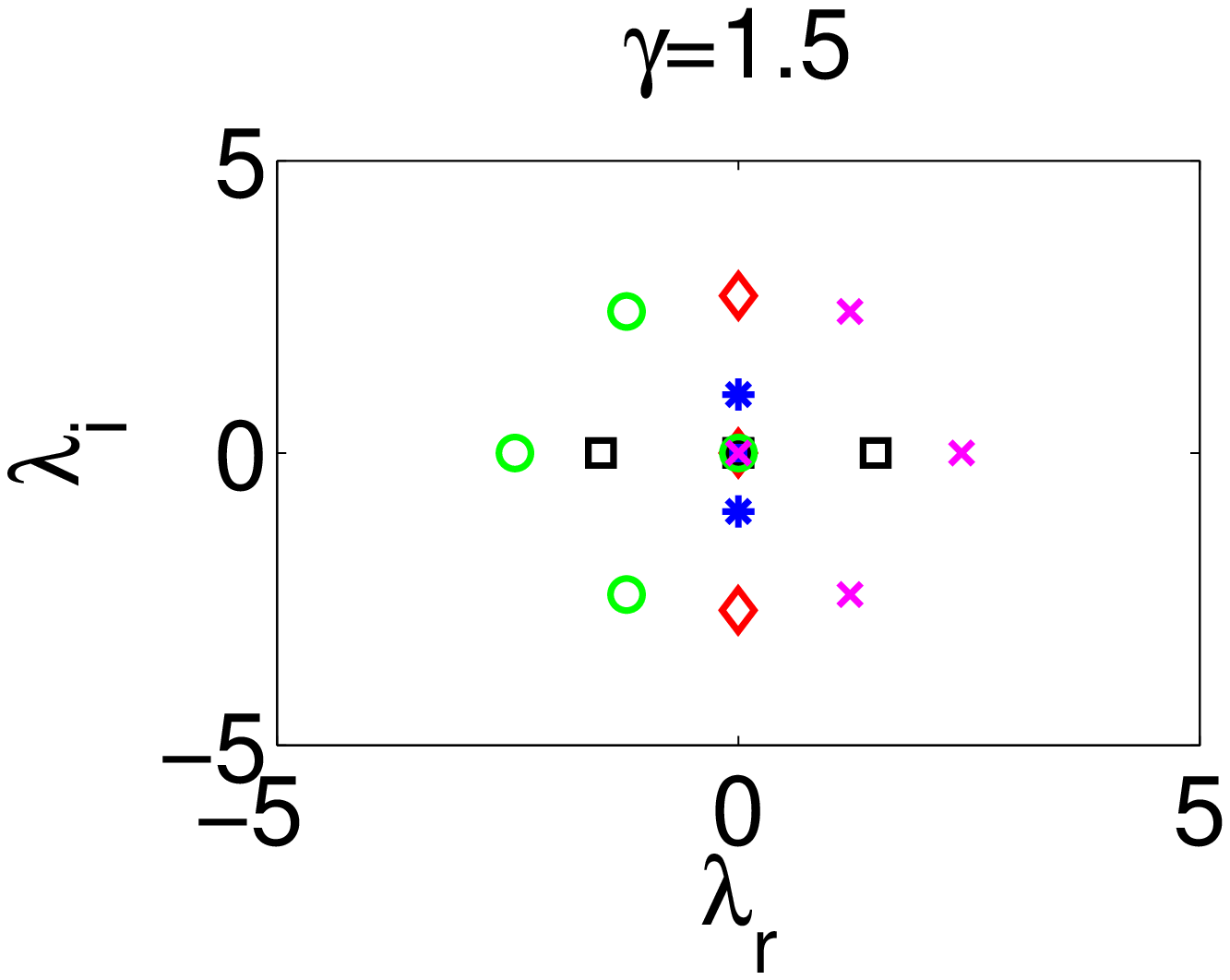}}
\caption{The eigenvalue plots illustrating the
linear stability of the nonlinear-PT-symmetric dimer with
$\epsilon=1$, $\rho_r=-2$ and $\rho_{im}=1$. For the blues stars and red diamonds branches, we use $E=1$ here, while for the case II (green circles
and magenta crosses) and case III (black squares), $E$ is determined
from the remaining parameters based on Eqs.~(\ref{en_condition})
and~(\ref{energy_eq}), respectively.}
\label{fig5}
\end{figure}

    Fig.~\ref{fig4} shows the profile of the different branches for
the dimer case and for parameters $\epsilon=1$, $E=1$, $\rho_r=-2$ and
$\rho_{im}=1$ (unless noted otherwise).
    The branches denoted by blue stars and red diamonds correspond to the
case I of symmetric solutions; these two branches collide and disappear
at the critical point $\gamma=1.61$
(when Eq.~(\ref{dimer1_restriction}) becomes an equality).
The green circle and magenta cross branches correspond
    to case II; and the black squares branch corresponds to case III.
For the latter two branches, when $\gamma$ is varied, $E$ is also varied
too (rather than staying fixed at $E=1$ as for case I)
according to Eqs.~(\ref{en_condition}) and~(\ref{energy_eq}), respectively.
Similar notations are used in Fig.~\ref{fig5}, which shows the linear
stability eigenvalues $\lambda = \lambda_r + i\lambda_{i}$ of the linearization.
While the branches of case I are stable, it is interesting
to note that  the branch of case III (black squares) is stable until a
pitchfork (symmetry breaking) bifurcation arises at $\gamma=0.895$
(when Eq.~(\ref{restrict_2}) becomes an equality) and
acquires a real pair of eigenvalues thereafter signalling its dynamical
instability. On the other hand, it is at that critical point that
the two branches belonging to case II arise. While the
special symmetric black squares' branch of case III persists
up to the critical point of
$\gamma=2 \epsilon=2$ of Eq.~(\ref{sine_relation}), it
should be pointed out that nonlinearity enables the asymmetric branches
of Case II to persist for large values of $\gamma$, in fact well
past the point of the linear PT phase transition.
This feature has been highlighted in a number of recent
works~\cite{pgk,konorecent3}; in the case of the dimer the linear
critical PT phase transition point is identified as $\gamma=\epsilon$,
while for the trimer setting considered below it is $\gamma=\sqrt{2} 
\epsilon$.

\begin{figure}[htp]
\subfigure[\ blue stars branch]{\scalebox{0.38}{\includegraphics{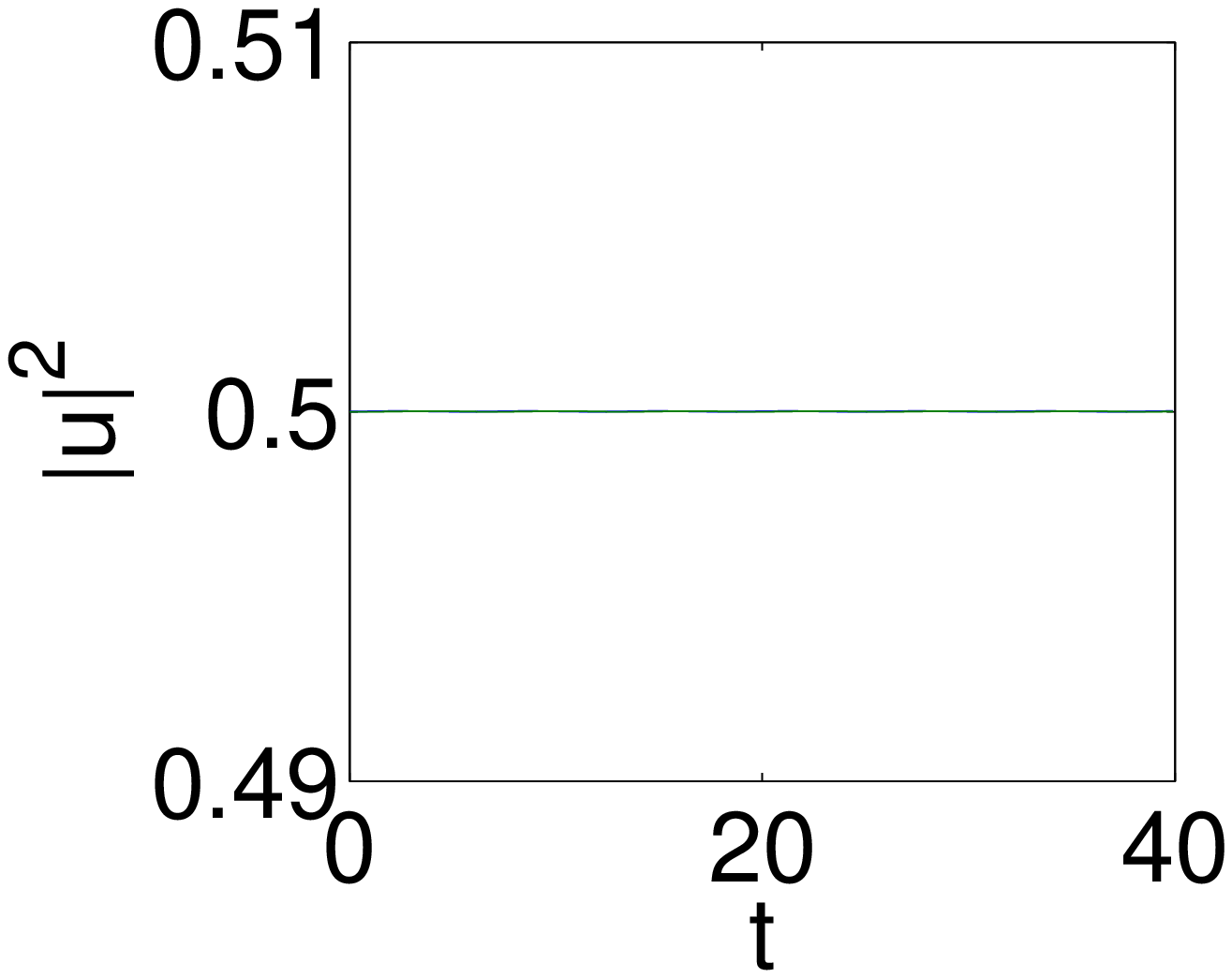}}}
\subfigure[\ red diamonds branch]{\scalebox{0.38}{\includegraphics{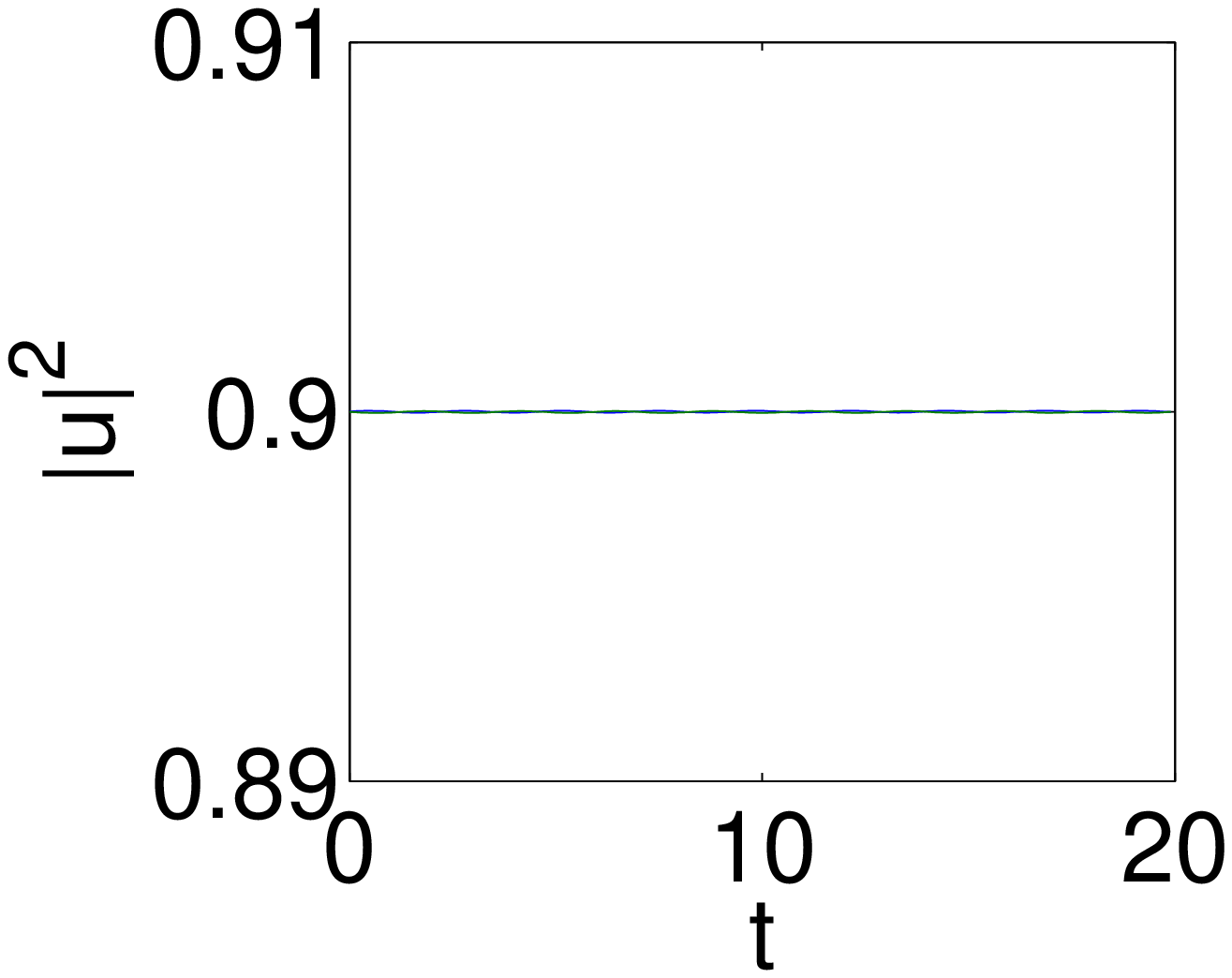}}}
\subfigure[\ black squares branch]{\scalebox{0.38}{\includegraphics{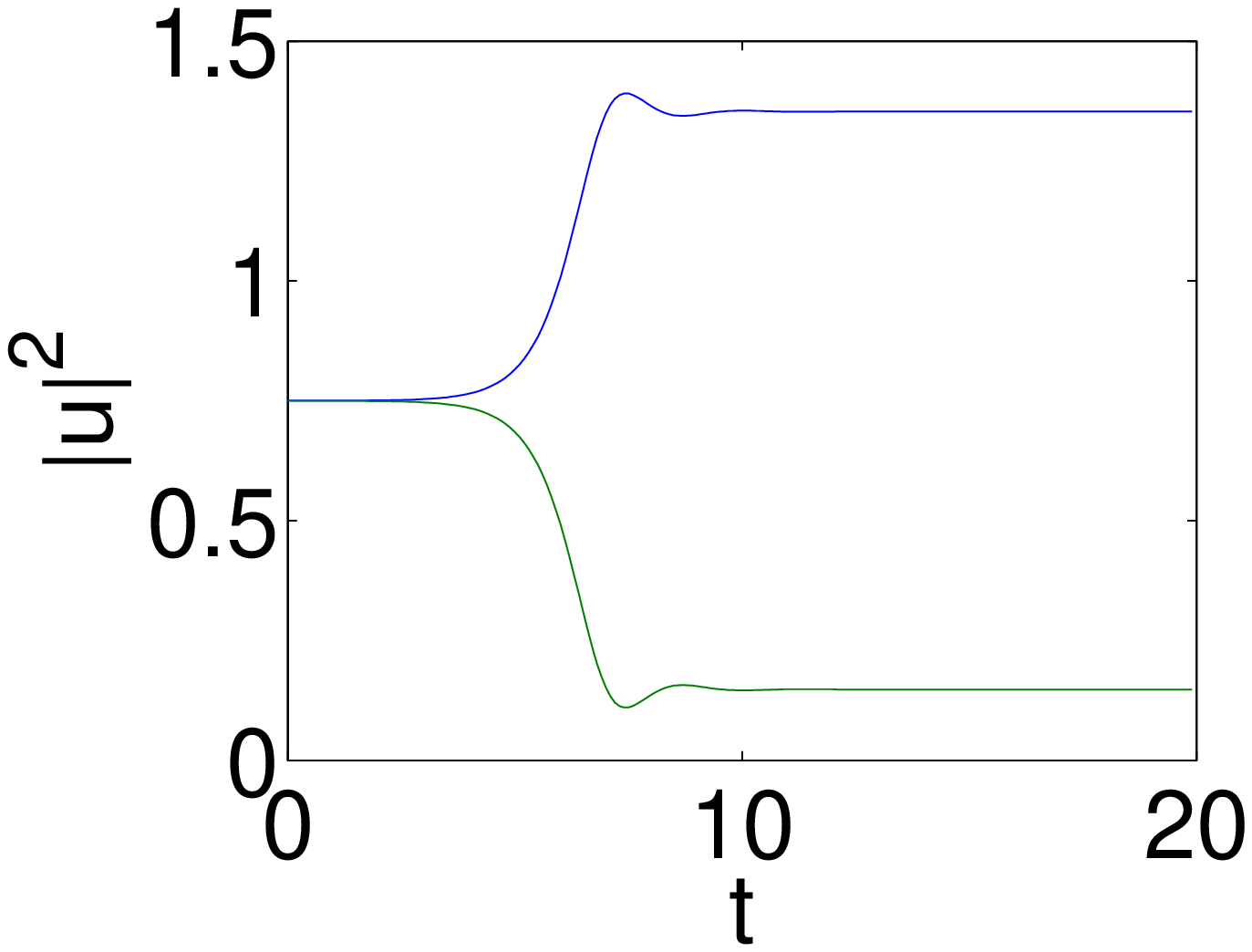}}}
\subfigure[\ green circles branch]{\scalebox{0.38}{\includegraphics{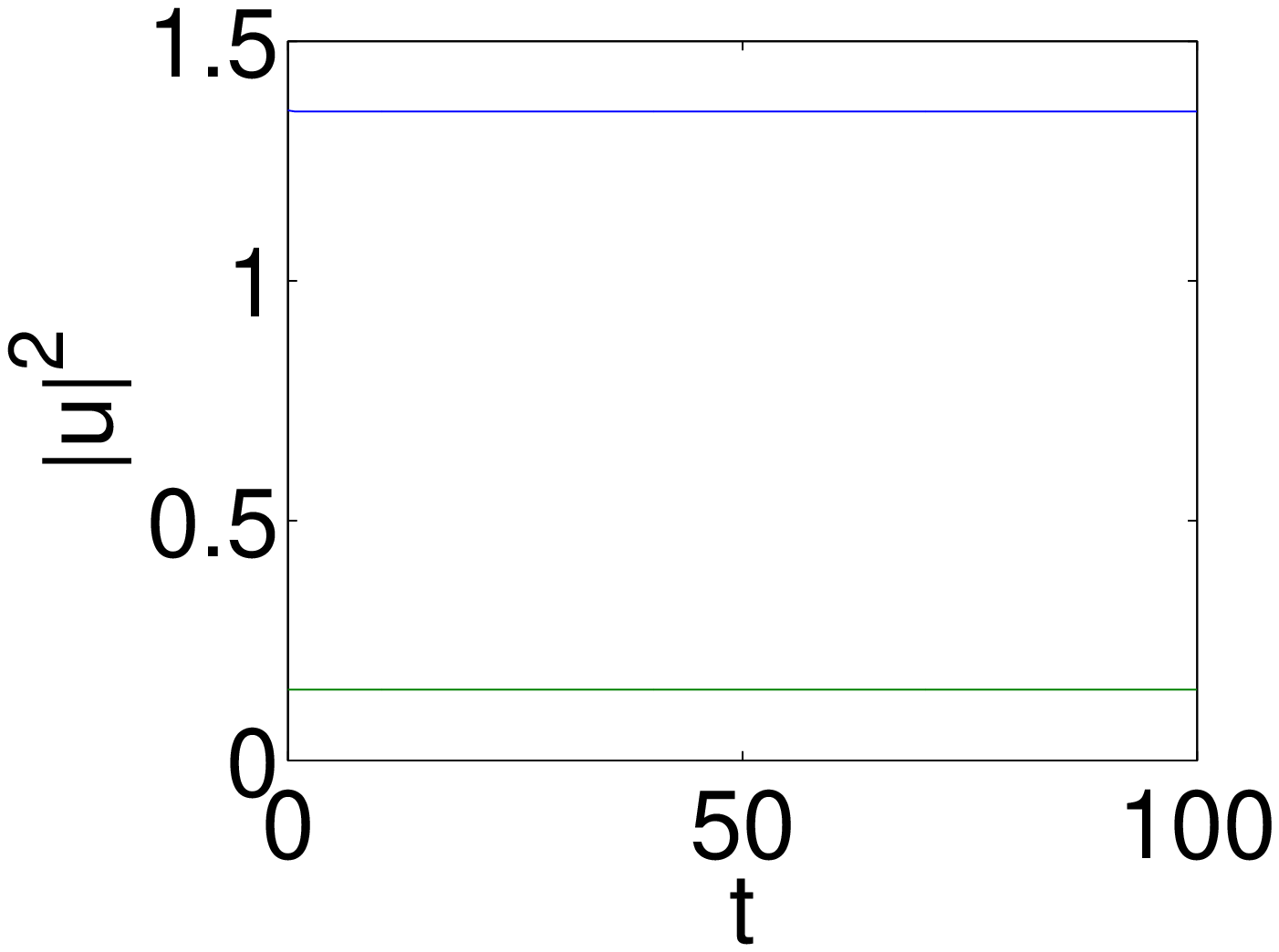}}}
\subfigure[\ magenta crosses branch]{\scalebox{0.38}{\includegraphics{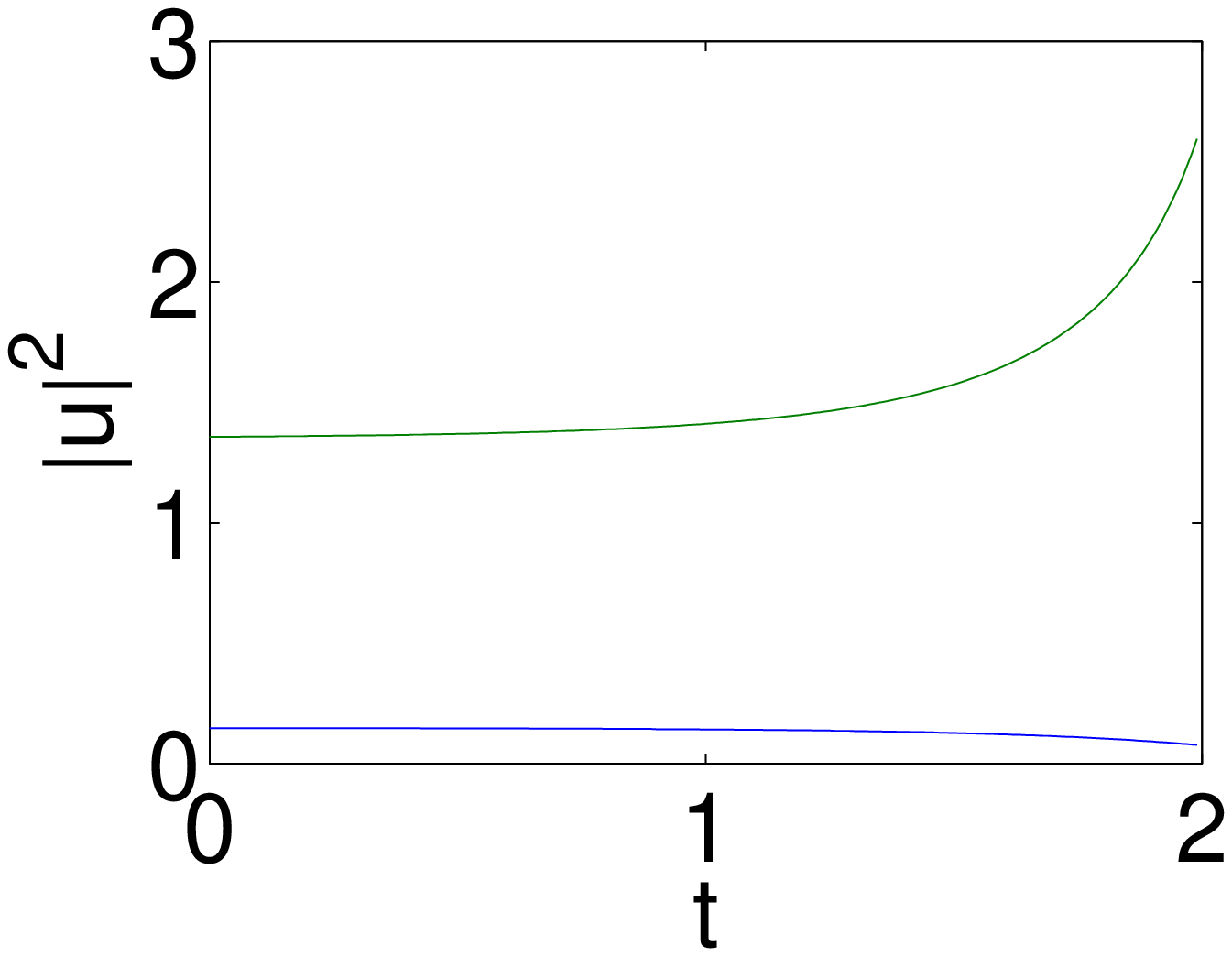}}}
\caption{The dynamical evolution plots of the branches for
the case of the nonlinear-PT-symmetric dimer with
the same parameter settings as in Fig.~\ref{fig5} when $\gamma=1.5$.
The symmetric blue stars and red diamonds of Case I and the asymmetric
green circles of Case II are stable, while the
black squares of Case III (past the pitchfork point) and
magenta crosses of Case II are unstable and deviate from their initial
profile during the dynamics (see also the discussion in the text).}
\label{fig6}
\end{figure}

    An additional point worthy to mention here is that
in linear PT-symmetric chains (just as is the case in typical
Hamiltonian systems), if $\lambda$ is an eigenvalue to the linearization
problem around a solution, so are $-\lambda$, $\bar{\lambda}$, and
$-\bar{\lambda}$ (where the overbar denotes complex conjugation here).
However, in our nonlinear PT-symmetric dimer $-\lambda$ and $-\bar{\lambda}$
may not appear in the linearization around a particular branch, as
is observed in Fig.~\ref{fig5}. Eigenvalues of the green circles and magenta
cross branches are not symmetric about the imaginary axis, but are
symmetric with respect to each other. One can see from Fig.~\ref{fig6} that
the green circles branch is always stable,
while the magenta crosses branch is always unstable (due to an oscillatory
instability associated with a complex eigenvalue pair).
This is because the existence of asymmetry in these solutions of case II
creates, in turn, asymmetries in the linearization matrix, due to the
nonlinear gain/loss term, which breaks the PT symmetry of the linearization
matrix and produces the corresponding observable asymmetry in eigenvalues.

The dynamical evolution of the different elements of the
bifurcation diagram of nonlinear-PT-symmetric
dimer is shown in Fig.~\ref{fig6} at a fixed
$\gamma=1.5$. In all the cases here and below, where a stationary
solution exists for the parameter value for which it is initialized,
a numerically exact solution up to $10^{-8}$ is typically used
as an initial condition in the system.  The system 
is sufficiently sensitive to dynamical 
instabilities that even the amplification of roundoff errors are enough to 
observe them.
The stability of the
case I branches is evident in the invariance of the relevant states during
the course of the simulation (blue stars and red diamonds). On the other
hand, the black squares branch is attracted towards the asymmetric
(yet stable, as is evident in the corresponding simulation)
green circles branch. Finally, the asymmetric magenta crosses branch
leads to indefinite growth of the site with the larger amplitude (nonlinear
gain) and the decay of the site with the smaller amplitude (nonlinear loss).

\section{Analysis of Stationary  Solutions for the Nonlinear-PT-Symmetric
Trimer Case}
We now consider the generalization of the above
considerations to the case of a so-called PT-symmetric trimer.
Here, the dynamical system associated with a potential application of
a three-waveguide setting is of the form:
\begin{eqnarray} \label{basic_trimer_equations}
iu_{t} = -\epsilon v + (\rho_{r} - i\rho_{im})|u|^{2}u + i \gamma u \nonumber \\
       \nonumber \\
iv_{t}= - \epsilon (u + w) - |v|^{2}v \hspace*{0.5in} \nonumber \\
       \nonumber \\
iw_{t} = - \epsilon v + (\rho_{r} + i\rho_{im})|w|^{2}w - i\gamma w
\end{eqnarray}
Such configurations have been considered earlier in optical applications
theoretically~\cite{majoh} and even experimentally~\cite{todd} in the
absence of gain/loss. Here in the spirit, of~\cite{kip}
(and also of~\cite{pgk}), we examine this case with both linear and nonlinear
gain/loss profiles. Once again, as in the case of the dimer, we present
the richer phenomenology setting of direct competition between linear
and nonlinear gain/loss. The middle site is assumed as devoid of gain
and loss. The Kerr nonlinearity is also assumed to be present in all
three sites.
Here, we use $u(t)$, $v(t)$
and $w(t)$ as the complex-valued components for the trimer.
For the stationary solutions, we again assume:
\begin{eqnarray} \label{complex_trimer_equations}
u(t) = ae^{iEt}, ~~~v(t) = be^{iEt} ~~~\mbox{and}~~~w(t) = ce^{iEt}.
\end{eqnarray}
Plugging Eqs.~($\ref{complex_trimer_equations}$) into Eq.~($\ref{basic_trimer_equations}$), we find:
\begin{eqnarray} \label{abc_trimer_nonlinear_equations}
  Ea = \epsilon b - (\rho_{r} - i\rho_{im})|a|^{2}a - i\gamma a, ~~~~~Eb = \epsilon (a + c) + |b|^{2}b \nonumber \\
               \nonumber \\
        Ec = \epsilon b - (\rho_{r} + i\rho_{im})|c|^{2}c + i\gamma c. \hspace*{0.75in}
\end{eqnarray}
Since $a, b$ and $c$ are complex-valued functions, we use the polar
decomposition:
\begin{eqnarray} \label{complex_abc_solutions}
 a = Ae^{i\phi_{a}}, ~~~b = Be^{i\phi_{b}} ~~~\mbox{and}~~~c = Ce^{i\phi_{c}}
\end{eqnarray}
where $A, B, C, \phi_{a}, \phi_{b}$ and $\phi_{c}$ are real-valued.
Plugging Eq.~($\ref{complex_abc_solutions}$) into Eq.~($\ref{abc_trimer_nonlinear_equations}$) and separating the
real and imaginary parts, we derive the following set of
real-valued equations for $A, B$ and $C$:
\begin{eqnarray}\label{trimer_ABC_nonlinear_equations}
  EA = \epsilon B \cos(\phi_{b} - \phi_{a}) - \rho_{r}A^{3}, ~~~~\epsilon B \sin(\phi_{b} - \phi_{a})
        + \rho_{im}A^{3} - \gamma A = 0 \hspace*{0.5in} \nonumber  \\
     \nonumber \\
 EB = \epsilon A \cos(\phi_{b} - \phi_{a}) + \epsilon C \cos(\phi_{b} - \phi_{c}) + B^{3}, ~~~~ - \epsilon A \sin(\phi_{b} - \phi_{a})
   - \epsilon C \sin(\phi_{b} - \phi_{c}) = 0 \nonumber \\
      \nonumber \\
  EC = \epsilon B \cos(\phi_{b} - \phi_{c}) - \rho_{r}C^{3}, ~~~~\epsilon B \sin(\phi_{b} - \phi_{c}) - \rho_{im}C^{3} + \gamma C = 0.
    \hspace*{0.5in}
\end{eqnarray}
We seek nontrivial solutions to Eqs.~($\ref{trimer_ABC_nonlinear_equations}$), i..e, $(A,B,C)\neq(0,0,0)$. We
can reduce Eqs.~($\ref{trimer_ABC_nonlinear_equations}$) to the form:
\begin{eqnarray} \label{reduced_trimer_nonlinear_eqns}
  \cos(\phi_{b} - \phi_{a}) = \frac{EA + \rho_{r}A^{3}}{\epsilon B}, ~~~~\cos(\phi_{b} - \phi_{c}) = \frac{EC + \rho_{r}C^{3}}{\epsilon B}
    \nonumber \hspace*{0.90in}\\
         \nonumber \\
  \sin(\phi_{b} - \phi_{a}) = \frac{\gamma A - \rho_{im}A^{3}}{\epsilon B}, ~~~~\sin(\phi_{b} - \phi_{c}) =
    -\frac{(\gamma C - \rho_{im}C^{3})}{\epsilon B} \nonumber \hspace*{0.65in} \\
        \nonumber \\
    A\cos(\phi_{b} - \phi_{a}) + C\cos(\phi_{b} - \phi_{c}) = \frac{EB - B^{3}}{\epsilon}, ~~~~
    A\sin(\phi_{b} - \phi_{a}) + C\sin(\phi_{b} - \phi_{c}) = 0.
\end{eqnarray}
We apply the first four equations of
Eqs.~($\ref{reduced_trimer_nonlinear_eqns}$) into the last two equations
and obtain the following relations:
\begin{eqnarray} \label{trimer_nonlinear_equations2}
   B^{4} - EB^{2} + E(A^{2} + C^{2}) + \rho_{r}(A^{4} + C^{4}) = 0, ~~~~
   (A^{2} - C^{2})[\gamma - \rho_{im}(A^{2} + C^{2})] = 0.
\end{eqnarray}
We now determine $A, B$ and $C$ for several subcases (symmetric, asymmetric
and mixed) as was done for the dimer case in section II.

\subsection{Existence of Localized Modes for the Trimer Case}
For the trimer case, the special cases that
can be seen to emerge for the solutions of
Eqs.~($\ref{trimer_nonlinear_equations2}$) can be classified as follows:
\begin{itemize}
   \item $\underline{ \mbox{Case I}:~~A^{2} = C^{2}  ~~\mbox{and}~~  A^{2} + C^{2} \neq \gamma/\rho_{im}}$:\\

             In this case the algebraic equations assume the form:
               \begin{eqnarray}
                   \cos(\phi_{b} - \phi_{a}) &=& \frac{EA + \rho_{r}A^{3}}{\epsilon B} = \cos(\phi_{b} - \phi_{c}) \nonumber \\
                       \nonumber \\
                   \sin(\phi_{b} - \phi_{a}) &=& \frac{\gamma A - \rho_{im}A^{3}}{\epsilon B} = -\sin(\phi_{b} - \phi_{c}).
               \end{eqnarray}
               We now
use Eq.~($\ref{trimer_nonlinear_equations2}$) and
                $\cos^{2}(\phi_{b} - \phi_{a}) + \sin^{2}(\phi_{b} - \phi_{a}) = 1$ to determine:
                 \begin{eqnarray} \label{trimer_subcase_eqns_A_equal_C_one}
                     (\rho_{r}^{2} + \rho_{im}^{2})A^{6} + 2(E\rho_{r} - \gamma \rho_{im})A^{4}
                       + (E^{2} + \gamma^{2})A^{2} - \epsilon^{2}B^{2} = 0, ~~~~~
                        B^{4} - EB^{2} + 2EA^{2} + 2\rho_{r}A^{4} = 0. \hspace*{0.5in}
                    \end{eqnarray}
                  One can solve eqns.~($\ref{trimer_subcase_eqns_A_equal_C_one}$)
                  for $A^{2}$ and $B^{2}$ to complete the calculation of
the relevant symmetric branch of solutions of Case I.
%
   \item $\underline{\mbox{Case II}:~~A^{2} + C^{2} = \gamma/\rho_{im} ~~\mbox{and}~~ A^{2} \neq C^{2}}$:
   \\

   From Eq.~(\ref{reduced_trimer_nonlinear_eqns}), we obtain the
four algebraic equations:
    \begin{eqnarray}\label{trimercase2}
       A^{2}(E+\rho_{\gamma}A^{2})^{2}+A^{2}(\gamma-\rho_{im}A^{2})^{2}=\epsilon^{2}B^{2} \nonumber \\
       \nonumber \\
       C^{2}(E+\rho_{\gamma}C^{2})^{2}+C^{2}(\gamma-\rho_{im}C^{2})^{2}=\epsilon^{2}B^{2} \nonumber \\
       \nonumber \\
       A^{2} + C^{2} = \frac{\gamma}{\rho_{im}} \nonumber \\
       \nonumber \\
       B^{4} - EB^{2} + E\left( A^{2}+C^{2}\right) + \rho_{r}\left( A^{4}+C^{4}\right)  = 0.
    \end{eqnarray}
        We now have four equations but with only three unknowns ($A$, $B$
and $C$). Therefore, in contrast to the previous symmetric branch
of case I, one of the
parameters $E,\ \epsilon,\ \rho_r,\ \rho_{im},\ \gamma$ is determined by the other four; i.e., not all four of these parameters can be picked independently
in order to give rise to a solution of the trimer. Once again, we should
nevertheless, highlight here that these asymmetric solutions only exist
because of the interplay of linear gain/loss and nonlinear loss/gain profiles.

      \item $\underline{\mbox{Case III}:~~A^{2} + C^{2} = \gamma/\rho_{im} ~~\mbox{and}~~ A^{2} = C^{2}}$: \\

     In this mixed case, we have
      \begin{align}
         A^{2}& = C^{2} = \frac{\gamma}{2\rho_{im}} \nonumber \\
         \nonumber \\
         B^{4}& - EB^{2} + 2EA^{2} + 2\rho_{r}A^{4} = 0,
      \end{align}
        with the restriction that
        \begin{eqnarray}
          E^{2}-\dfrac{4E\gamma}{\rho_{im}}-\dfrac{2\rho_{r}\gamma^{2}}{\rho_{im}^{2}}\ge0.
        \end{eqnarray}
        One can solve the following equations for $B$ and $E$
        \begin{eqnarray}\label{trimercase3}
         B^{4} - EB^{2} + 2E\dfrac{\gamma}{2\rho_{im}} + 2\rho_{\gamma}\dfrac{\gamma^{2}}{4\rho_{im}^2} &=& 0 \nonumber \\
         \nonumber \\
         \left( \rho_{\gamma}^{2}+\rho_{im}^{2}\right) \dfrac{\gamma^{3}}{8\rho_{im}^{3}}+(2\rho_{im}E-2\rho_{im}\gamma)\dfrac{\gamma^{2}}{4\rho_{im}^{2}}+\left( E^{2}+\gamma^{2}\right) \dfrac{\gamma}{2\rho_{im}}&=&\epsilon^{2}B^{2}.
         \end{eqnarray}
         These equations imply that one of the parameters (e.g., $E$) will be determined once the parameters, $\gamma,\ \epsilon,\ \rho_r$ and $\rho_{im}$ are chosen.
\end{itemize}

 \subsection{Linear Stability Analysis for the Trimer Case}

  We again consider the nonlinear PT-symmetric trimer
model with linear and nonlinear gain/loss and examine the linear stability of its
solutions given in Eq.~(\ref{basic_trimer_equations}) for the solutions
  given in the previous section. We begin by positing the linearization
ansatz:
  \begin{eqnarray} \label{uvw_stability_solutions}
      u(t) = e^{iEt}[ a + pe^{\lambda t} + Pe^{\lambda^{*}t}], ~~~~~v(t) = e^{iEt}[ b + qe^{\lambda t} + Qe^{\lambda^{*}t}] \nonumber \\
                 \nonumber \\
            w(t) = e^{iEt}[ c + re^{\lambda t} + Re^{\lambda^{*}t}] \hspace*{1.0in}
  \end{eqnarray}
where $p, P, q, Q, r, R$ are perturbations to the solutions of interest.
Plugging
  Eq.~($\ref{uvw_stability_solutions}$) into Eq.~(\ref{basic_trimer_equations}) and truncating at the linear order
  in $p, P, q, Q, r$ and $R$, we derive the following eigenvalue problem:
    \begin{eqnarray}
       {\bf A}{\bf Y} = i\lambda {\bf Y}
    \end{eqnarray}
    where ${\bf Y} = (p, ~-P^{*}, ~q, ~-Q^{*}, ~r, ~-R^{*})^{T}$ and ${\bf A}$ is the $(6 \times 6)$ matrix:
    \begin{eqnarray}
       {\bf A} = \left( \begin{array}{cccccc}
                                    a_{11} & -a^{2}(\rho_{r} - i\rho_{im}) & -\epsilon & 0 & 0 &0\\
                                    (a^{*})^{2}(\rho_{r} + i\rho_{im}) & a_{22} & 0 & \epsilon & 0 & 0\\
                                    -\epsilon & 0 &  E - 2|b|^{2} &  b^{2} & -\epsilon & 0\\
                                     0 & \epsilon & -(b^{*})^{2} & -E + 2|b|^{2} & 0 &\epsilon\\
                                   0 &0 & -\epsilon &0 & a_{55} &  -c^{2}(\rho_{r} + i\rho_{im})\\
                                    0 & 0 & 0 & \epsilon & (c^{*})^{2}(\rho_{r} - i\rho_{im}) & a_{66}
                               \end{array}\right)
    \end{eqnarray}
    where
    \begin{eqnarray}
        a_{11} = E + i\gamma + 2|a|^{2}(\rho_{r} - i\rho_{im})
     \nonumber \\
            \nonumber \\
        a_{22} = -E + i\gamma - 2|a|^{2}(\rho_{r} + i\rho_{im}) \nonumber \\
            \nonumber \\
       ~~~a_{55} = E - i\gamma + 2|c|^{2}(\rho_{r} + i\rho_{im}), 
       \nonumber \\
          \nonumber \\
        ~~~a_{66} = -E - i\gamma - 2|c|^{2}(\rho_{r} - i\rho_{im}).
    \end{eqnarray}
The solution of this $6 \times 6$ eigenvalue problem (and whether
the corresponding eigenvalues $\lambda$ possess a positive real part)
will determine the spectral stability properties of the solutions
of the nonlinear-PT-symmetric trimer.

    \subsection{Numerical Results for the Trimer Case}
    \begin{itemize}
    \item Trimer Case I:

    The numerical results for the symmetric solutions of
the nonlinear-PT-symmetric trimer (Case I) are shown in Fig.~\ref{fig7},
Fig.~\ref{fig8} and Fig.~\ref{fig9}, with similar notations as in the dimer
case. Solutions are found by numerically solving
Eq.~(\ref{trimer_subcase_eqns_A_equal_C_one}).
A typical example of the branches that may arise in case I
of the trimer is shown for the parameters $\epsilon=1$, $E=1$, $\rho_r=-1$
and $\rho_{im}=1$. In this case, we find three branches in the considered
interval of parameter values. There are two branches which exist up to the
point $\gamma=2.59$ where they collide in a saddle-node bifurcation. One of
these, the red diamonds branch, is mostly unstable except for
$\gamma \in [1.26 ,1.33]\cup [2,2.11]$. For $\gamma<1.26$, this branch
has one real and one imaginary pair, which become both imaginary for
$\gamma>1.26$ until they collide for $\gamma=1.33$ and yield a complex
quartet, which subsequently splits into two imaginary pairs for $\gamma=2$
and finally into one real and one imaginary pair for $\gamma>2.11$.
    The other one, the black squares branch, is always unstable due
to one real and one imaginary pair. When these two modes collide,
a collision arises between both their real and their imaginary
(respective) eigenvalue pairs.

    Aside from the other two branches, the branch associated with the
blue stars emerges  from $\gamma=1$ and persists beyond the above
critical point (and for all values of $\gamma$ that we have monitored).
In our case, this branch is only stable for $\gamma<1.25$, at
which two pairs of imaginary eigenvalues collide and lead to a
complex quartet, which renders the branch unstable thereafter. This branch
behaves very similarly as the one in the linear PT trimer case
reported in~\cite{pgk}.
Both of them bifurcate from zero amplitude after a certain value of $\gamma$,
persist beyond the linear PT
critical point and have similar stability properties.

    \begin{figure}[htp]
    \scalebox{0.5}{\includegraphics{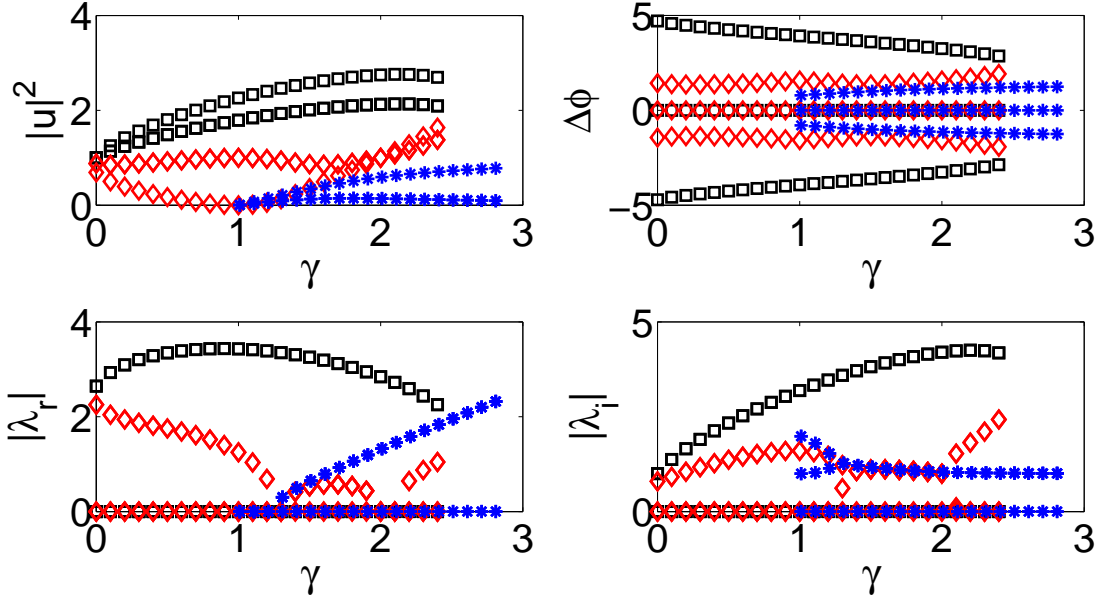}}
    \caption{The symmetric solution profiles of Case I in the
nonlinear-PT-symmetric trimer
with $\epsilon=1$, $E=1$, $\rho_r=-1$ and $\rho_{im}=1$.
    The three branches are denoted by blue stars, red diamonds and
black squares and their amplitudes (top left), phases (top right),
real part (bottom left) and imaginary part (bottom right) of the
corresponding eigenvalues are shown. See also the relevant discussion in the
text.}
    \label{fig7}
    \end{figure}

    \begin{figure}[htp]
    \scalebox{0.38}{\includegraphics{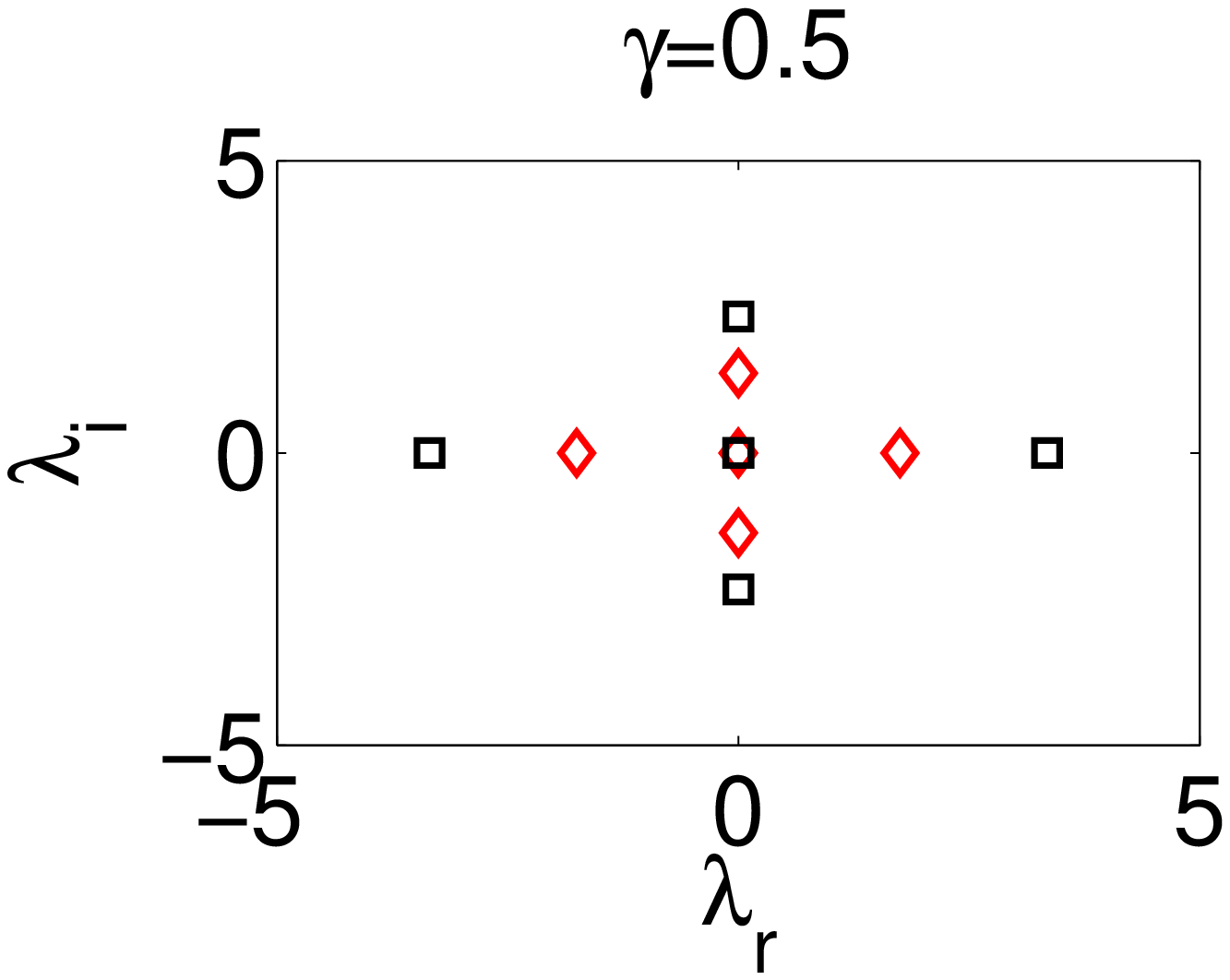}}
    \scalebox{0.38}{\includegraphics{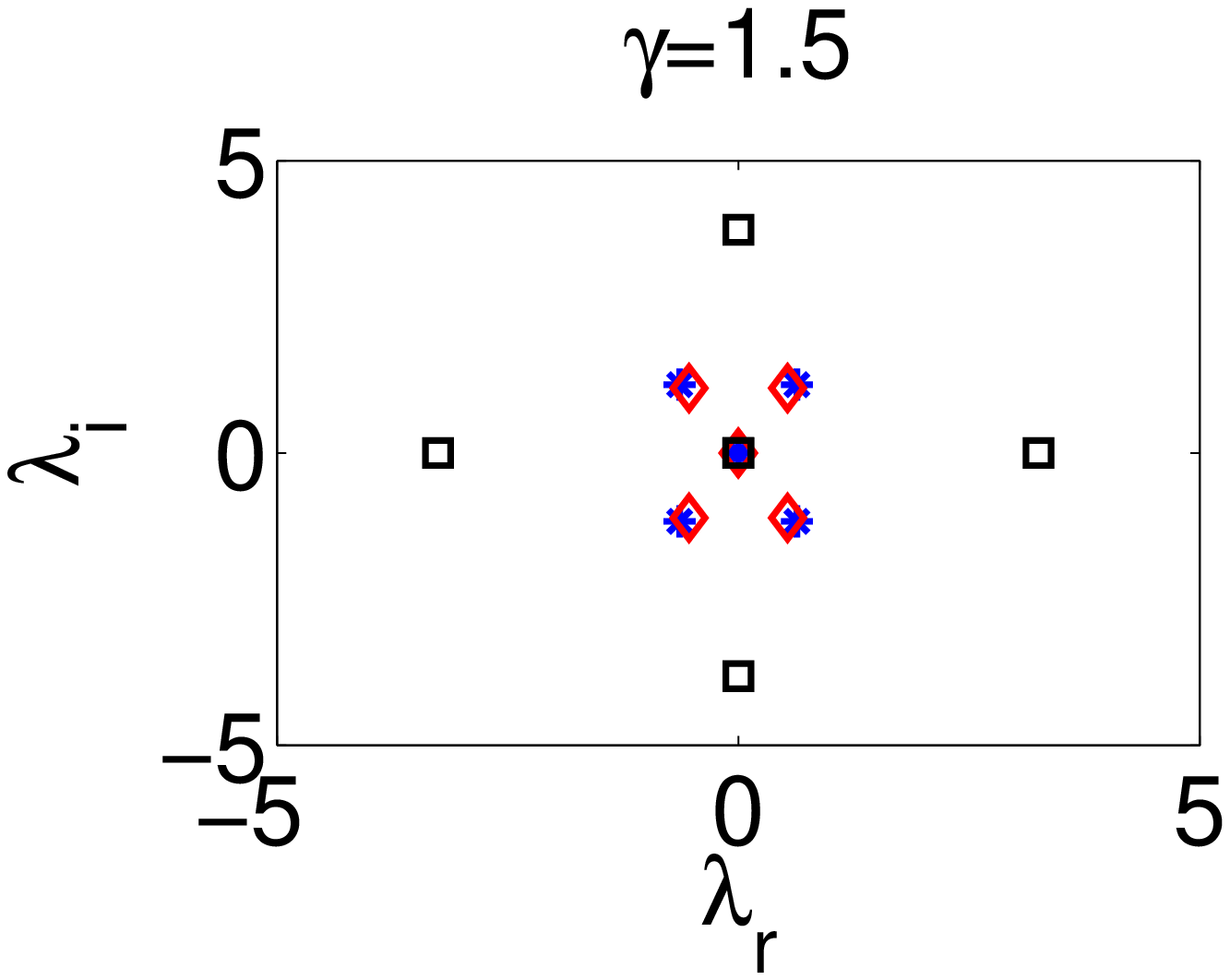}}
    \scalebox{0.38}{\includegraphics{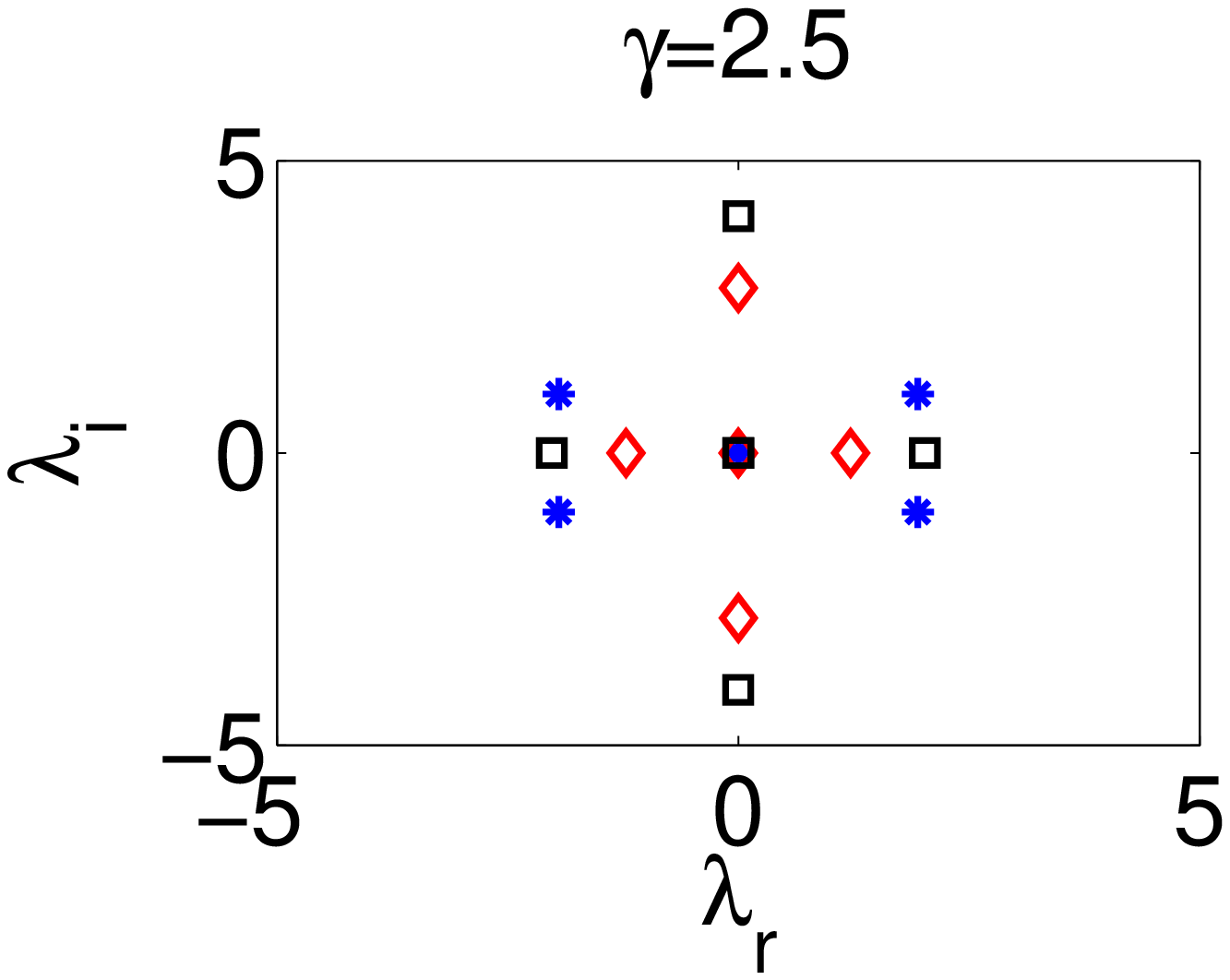}}
    \caption{The spectral plane of the linear stability analysis
for the symmetric solutions of Case I with
$\epsilon=1$, $E=1$, $\rho_r=-1$ and $\rho_{im}=1$, for three
different values of $\gamma=0.5$, $1.5$ and $2.5$. Each branch is
associated with three eigenvalue pairs one of which is at $0$ due
to symmetry.}
    \label{fig8}
    \end{figure}

To monitor the dynamical evolution of the different branches, we used
direct numerical simulations illustrated in Fig.~\ref{fig9} for
the case of $\gamma=1.5$. Two of the branches, the blue stars
of the left panel and the red diamonds of the middle one are oscillatorily
unstable for this value of $\gamma$, while the black squares branch
is always unstable due to a real eigenvalue pair. The latter has been
found to generically cause the unbounded gain of at least one node
within the trimer. The oscillatory instability, on the other hand,
in the case of the blue branch and for $\gamma=1.5$ can be observed
to lead to a long-lived periodic exchange of ``power'' between the
three sites. On the other hand, for the red diamonds branch of the
middle panel, while there is an intermediate stage of power oscillations
between the three nodes, the ultimate fate of the configuration favors
the unbounded growth of at least one node (in fact, two nodes
in the example shown) of the trimer.

    \begin{figure}[htp]
    \subfigure[\ blue stars branch]{\scalebox{0.38}{\includegraphics{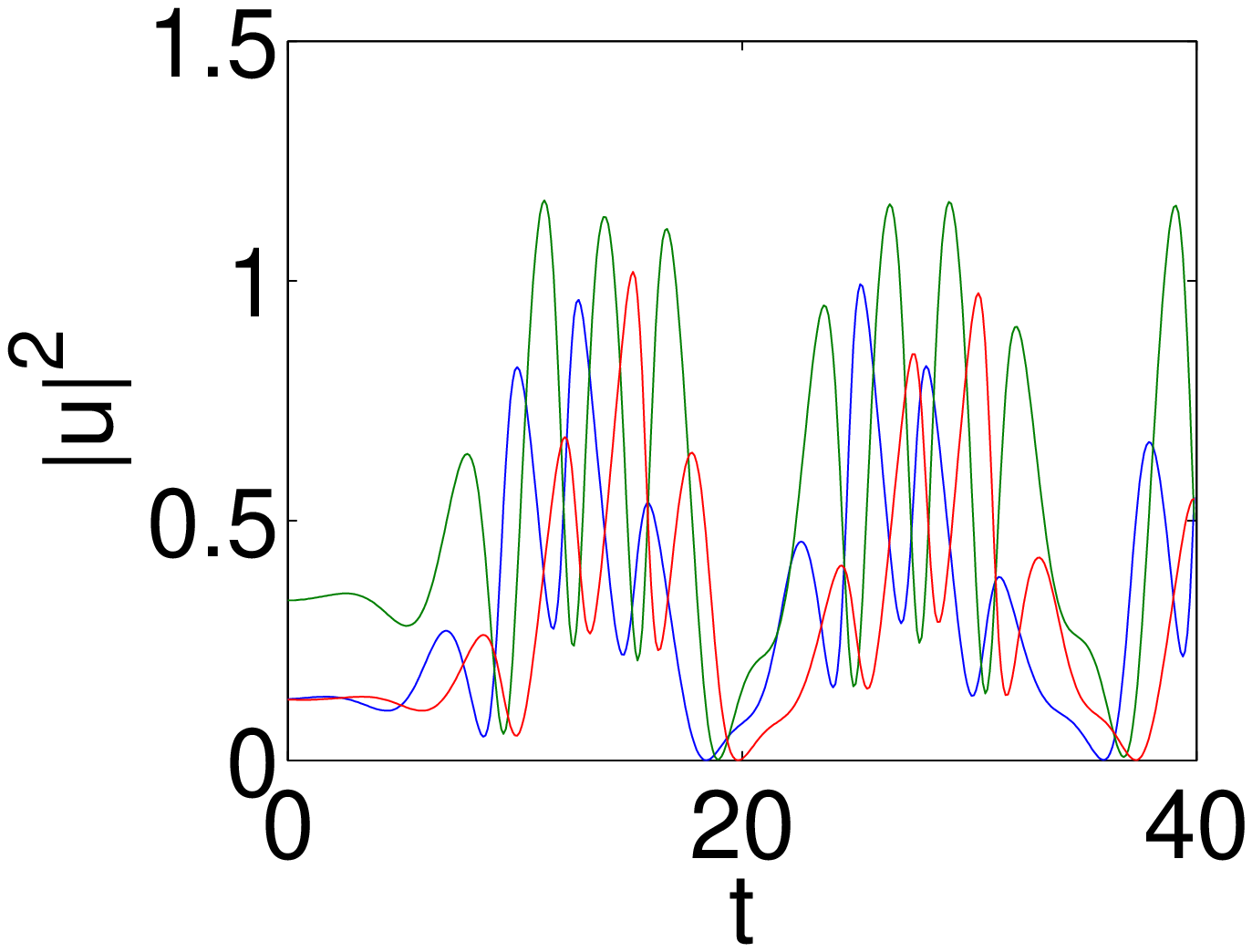}}}
    \subfigure[\ red diamonds branch]{\scalebox{0.38}{\includegraphics{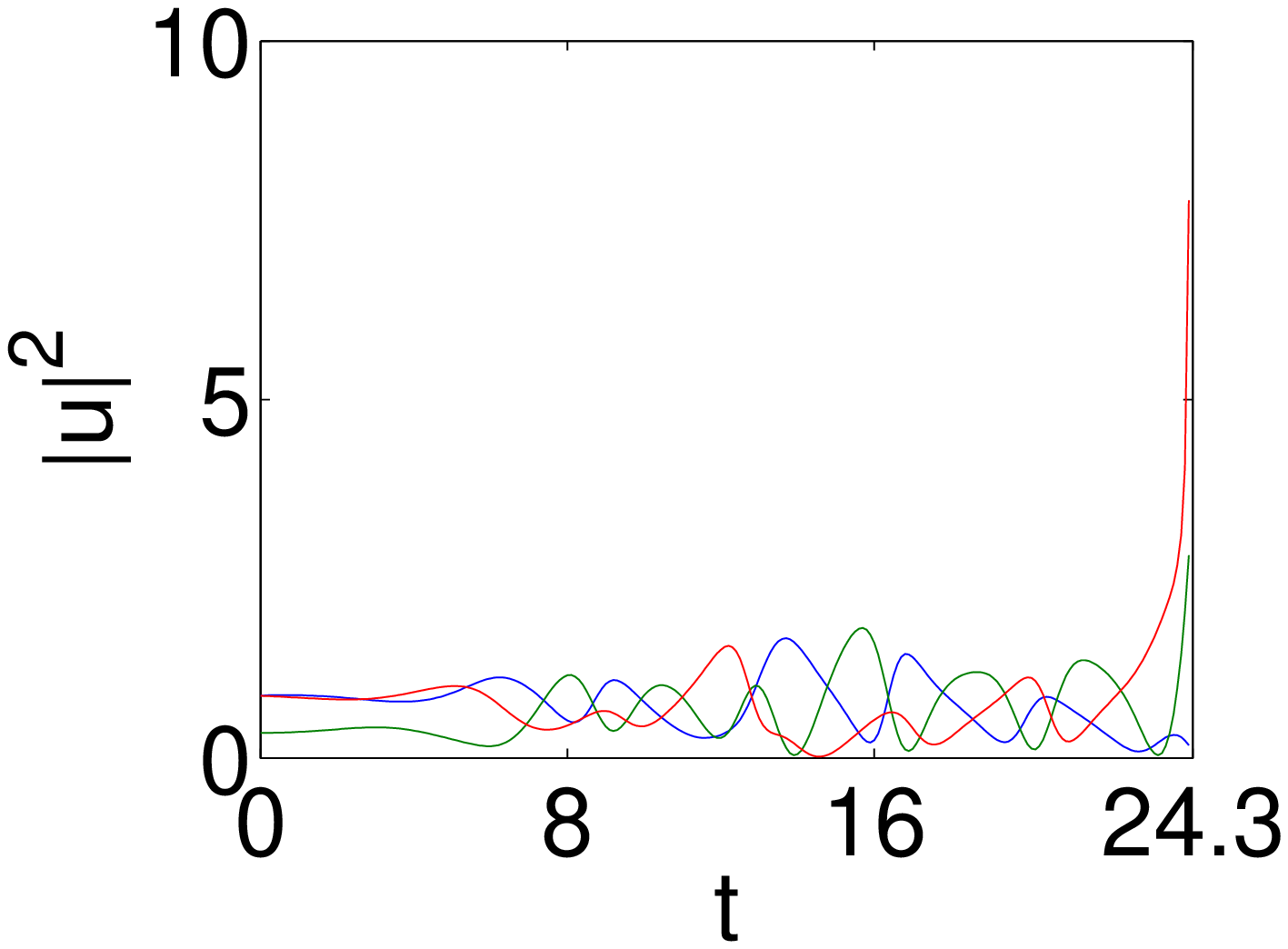}}}
    \subfigure[\ black squares branch]{\scalebox{0.38}{\includegraphics{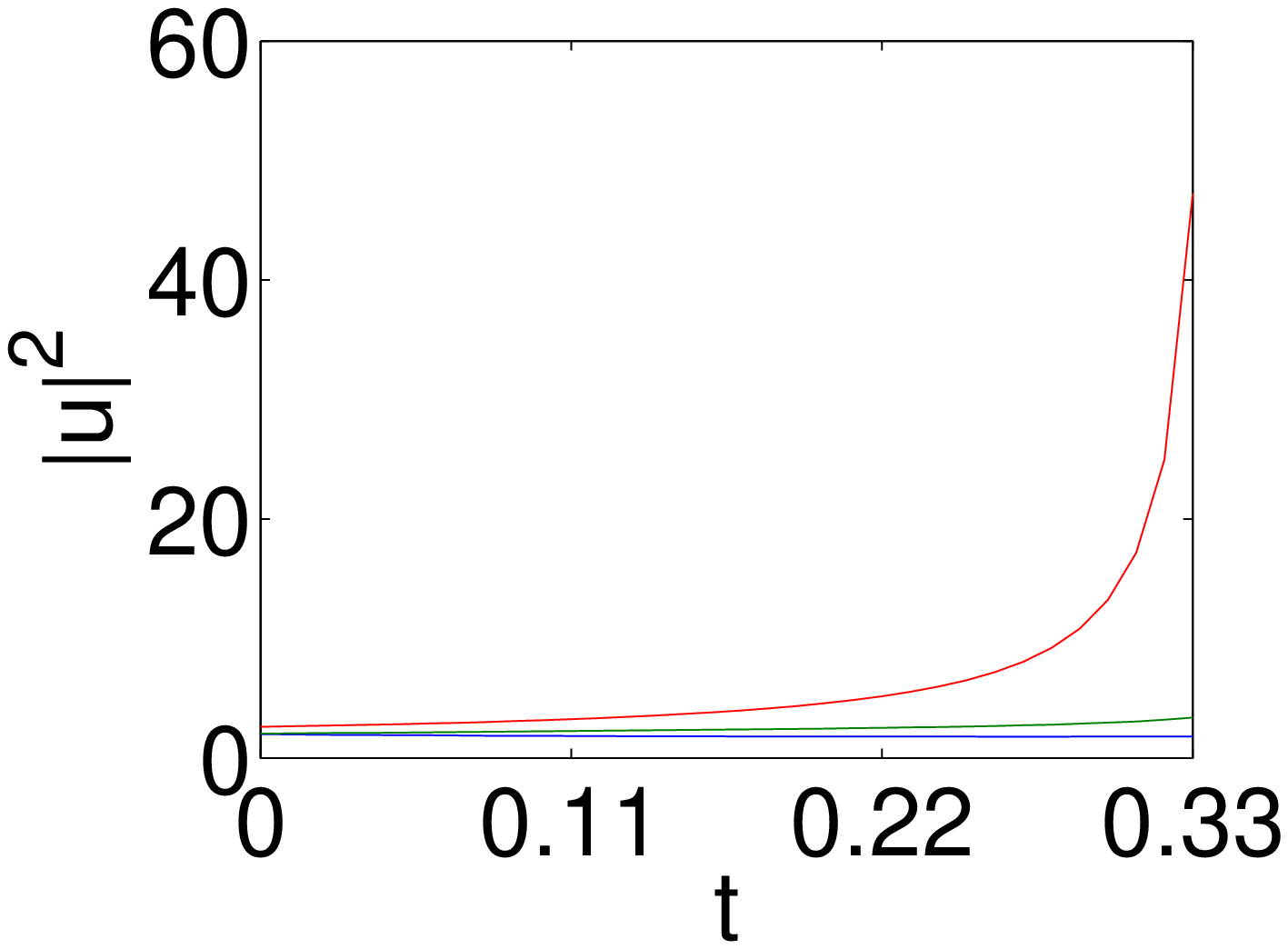}}}
    \caption{The time evolution plots of the trimer case I with $\epsilon=1$, $E=1$, $\rho_r=-1$ and $\rho_{im}=1$ when $\gamma=1.5$. For each branch, the blue
line denotes the nonlinear
loss/linear gain site, the
red line denotes the nonlinear gain/linear loss site, while the green represents the inert site between
the two. Notice the oscillatory evolution of the blue solid branch, while
the red diamonds and black squares lead to ultimate unbounded increase
of at least one site within the trimer.}
    \label{fig9}
    \end{figure}

    \item Trimer Case II:

    According to Eq.~(\ref{trimercase2}), one of the parameters should be determined by the others. We hereby set $\epsilon=1$, $\rho_r=-1$ and $\rho_{im}=1$, then $E$ is obtained self-consistenly for a given choice of $\gamma$. The
solution profiles, which are obtained by solving Eq.~(\ref{trimercase2})
numerically, are plotted in Fig.~\ref{fig10}. 
There are three pairs of branches, i.e., six branches of solutions in total
found in this case. Only one out of each pair is shown in Fig.~\ref{fig10}-\ref{fig12} (to avoid cluttering of the relevant figures), namely the blue stars, red diamonds and black squares. The other three branches are mirror symmetric to 
these branches, respectively. For example, the existence profile of the mirror 
symmetric branch of the blue stars would be identical to the blue stars 
shown in Fig.~\ref{fig10}, while its stability plot would be mirror symmetric 
about the imaginary axis to the blue stars shown in Fig.~\ref{fig11}.
Among the three branches shown in Fig.~\ref{fig10},
two of them emerge at $\gamma=2$ and persist throughout
the range of $\gamma$ values considered.
It should be noticed that the amplitudes are different within each
branch. The blue stars (dynamically stable) branch has a large $A$ and small
$B$ and $C$, while the red diamonds and black squares branches have a fairly small $A$ and large values of $B$ and $C$. In fact, precisely at the critical
point of the branches' emergence, the blue stars and the black squares
are exact mirror images of each other (i.e., they have the same amplitude
for $B$ and the one's $A$ is the other's $C$ -and vice versa-). This mirror
symmetry is in fact directly reflected in the eigenvalues of the linearization
around the two configurations, one set of which (for the blue stars)
possesses negative real parts, while the other (black squares) has mirror
symmetric positive ones. As can be perhaps intuitively anticipated, the
more stable configuration is the one having large amplitude at the nonlinear
loss/linear gain site. The third branch (red diamonds) is also highly
unstable and emerges out of a bifurcation at $\gamma=2.05$ (to which
we will return when discussing case III).

Fig.~\ref{fig11} shows the eigenvalues of the three branches clearly
illustrating the fact that they
are not symmetric about the imaginary axis. This can once again be
justified by the asymmetry of the configurations of case II which, in turn,
break the PT symmetry of the linearization matrix and hence lead to asymmetric
spectra.
The blue stars branch is always stable, as indicated above, and the other two
branches are always unstable as $\gamma$ increases. For instance,
for the red diamond branch, there exists (in addition to a zero eigenvalue)
an imaginary pair, a complex conjugate pair (with a positive
real part) and a real eigenvalue. In the case of the black squares branch,
there are (in addition to the zero eigenvalue) three real pairs (two positive
and one negative) and a complex conjugate pair (with positivre real part).

Fig.~\ref{fig12} shows the dynamical plots of the three distinct branches of solutions. 
The blue stars branch clearly preserves its configuration due to
its dynamical stability, while for
the two unstable branches, their evolution gives rise to asymmetric dynamics
favoring the loss of the power in a single site (the nonlinear gain/linear
loss one), and quickly absorbed by a stable state. The latter state
appears to be the mirror symmetric of the red diamonds asymmetric
branch in both cases. This is indeed also
a stable dynamical state of the original stationary system of equations.
Furthermore, this state is expected to exist
based on the symmetry breaking bifurcation
that we will discuss below as giving rise to the red diamonds branch.
As remarked above, the stability properties of  the 
mirror symmetric branches are mirror
symmetric to the ones shown in Fig.~\ref{fig11}. This implies
that only the mirror symmetric branch of the red diamonds is stable, which is,
in turn, consonant with our observation that it is a potential attractor
for the dynamics for $\gamma=3$ shown in Fig.~\ref{fig12}.

    \begin{figure}[htp]
    \scalebox{0.5}{\includegraphics{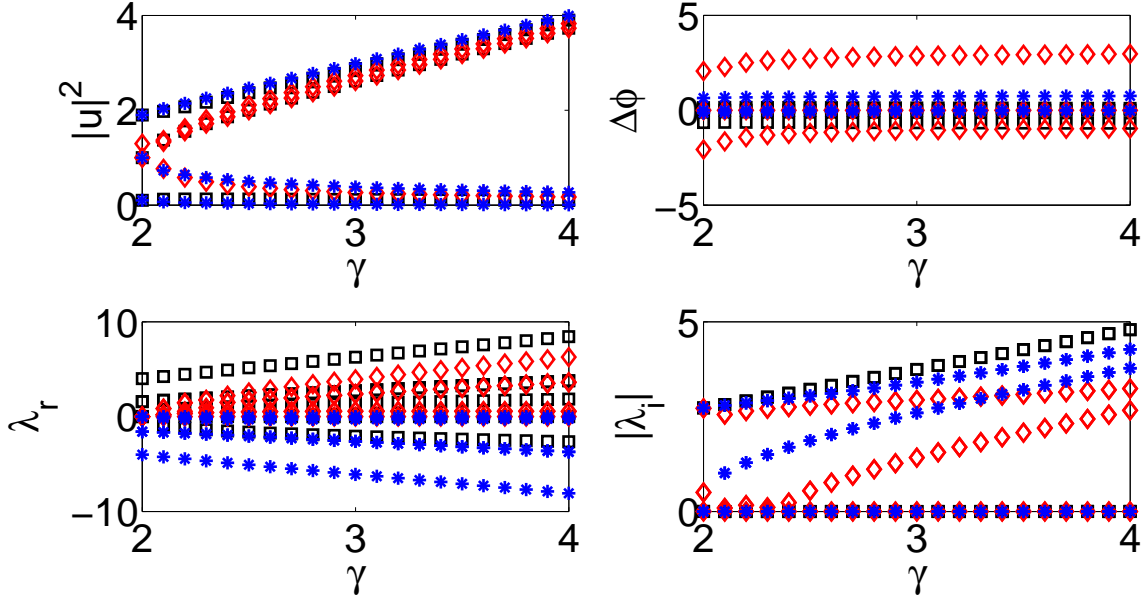}}
    \caption{The solution profile of the trimer case II with $\epsilon=1$, $\rho_r=-1$ and $\rho_{im}=1$.
    The three branches are denoted by blue stars, red diamonds and black squares. These branches start at $\gamma=2$ (except for the red diamonds branch
that is initiated at $\gamma=2.05$)
    and exist even when $\gamma$ is large.}
    \label{fig10}
    \end{figure}

    \begin{figure}[htp]
    \scalebox{0.4}{\includegraphics{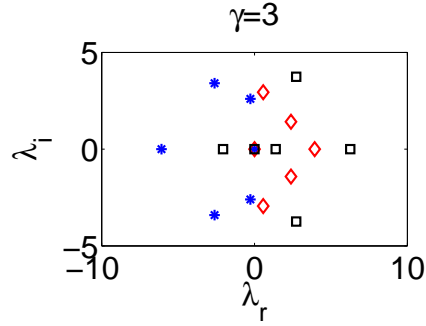}}
    \caption{The spectral stability plots of the trimer case II with
$\epsilon=1$, $\rho_r=-1$ and $\rho_{im}=1$, illustrating the stability
of the blue stars' branch and the instability of the other two.}
    \label{fig11}
    \end{figure}

    \begin{figure}[htp]
    \subfigure[\ blue stars branch]{\scalebox{0.38}{\includegraphics{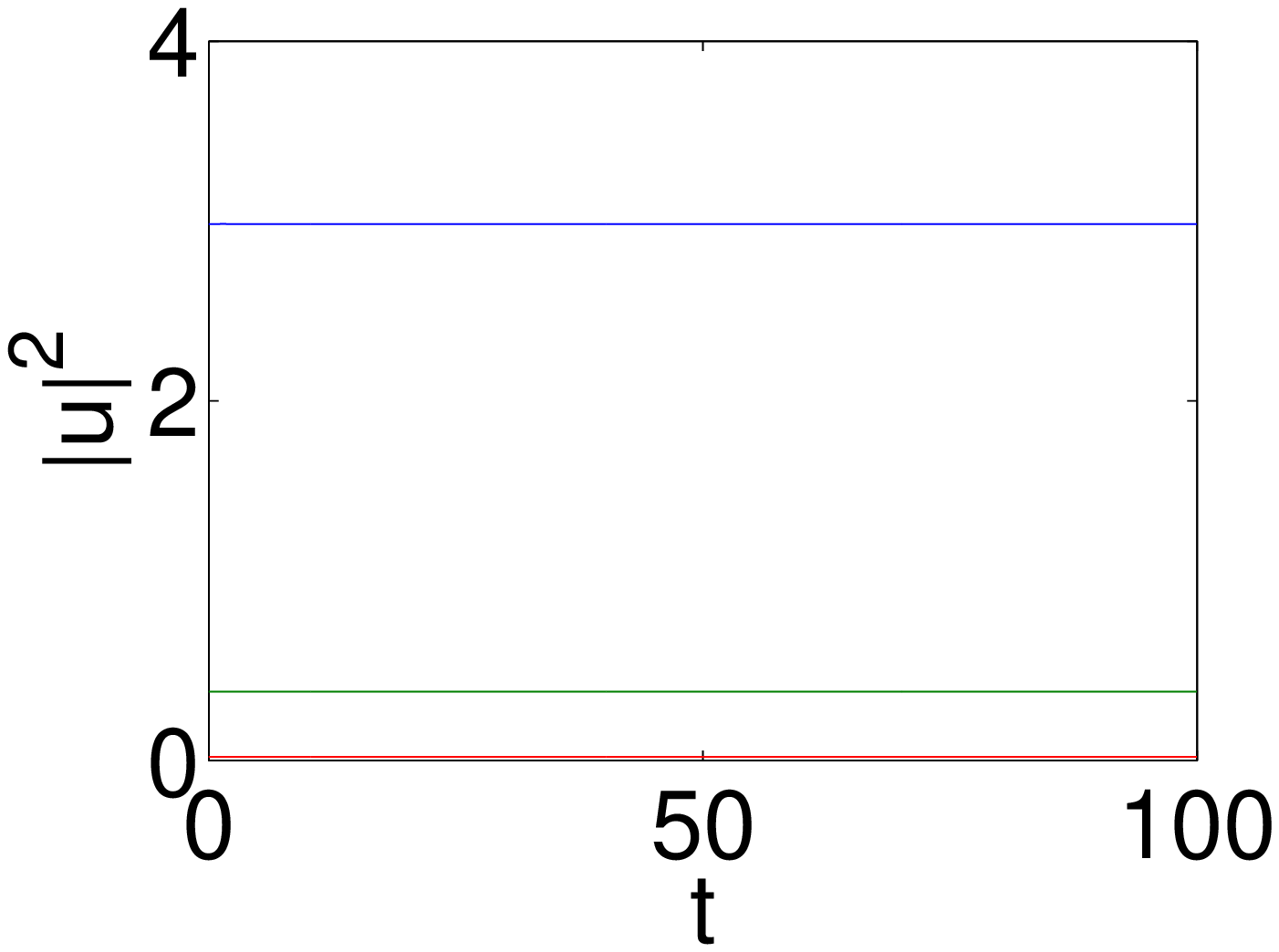}}}
    \subfigure[\ red diamonds branch]{\scalebox{0.38}{\includegraphics{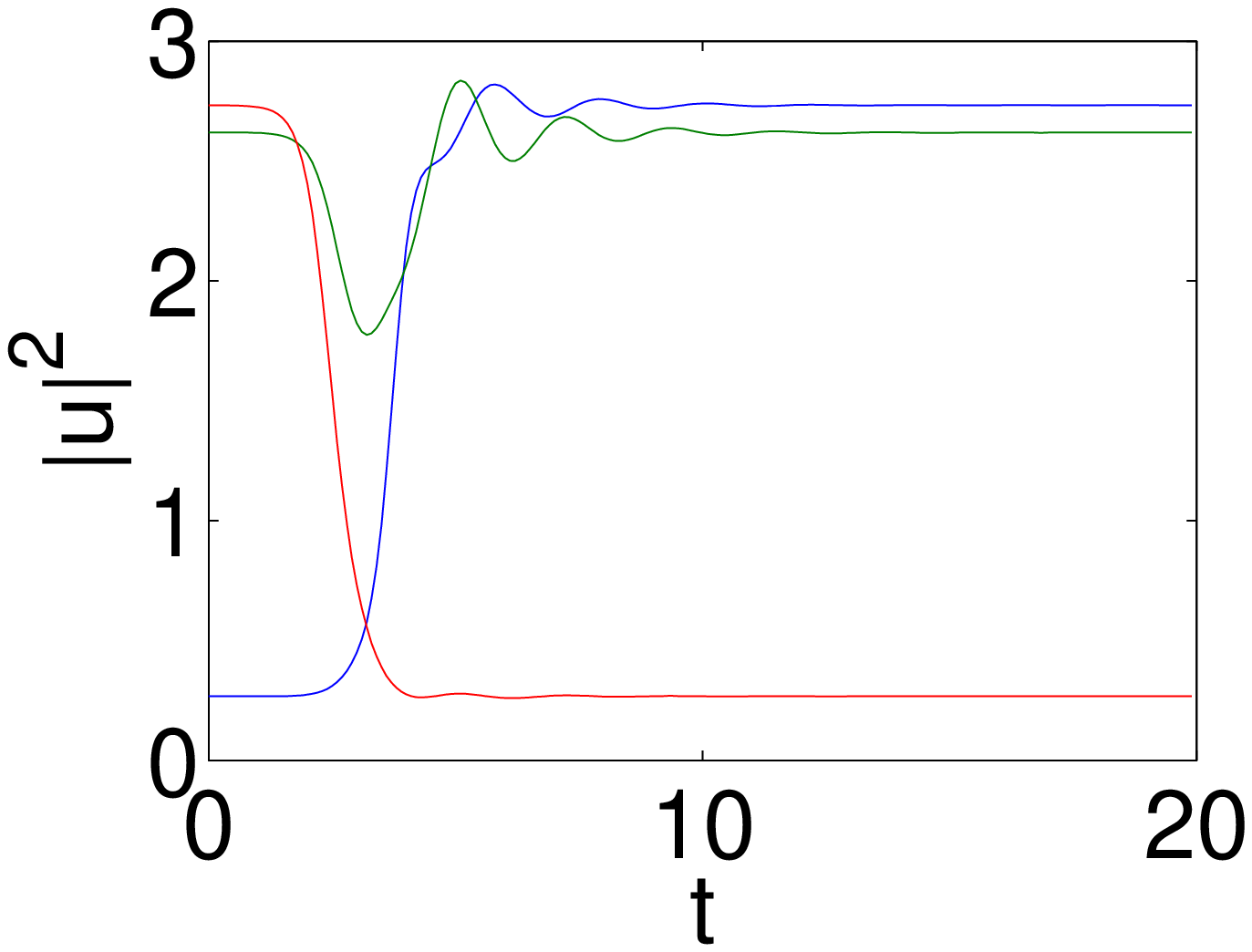}}}
    \subfigure[\ black squares branch]{\scalebox{0.38}{\includegraphics{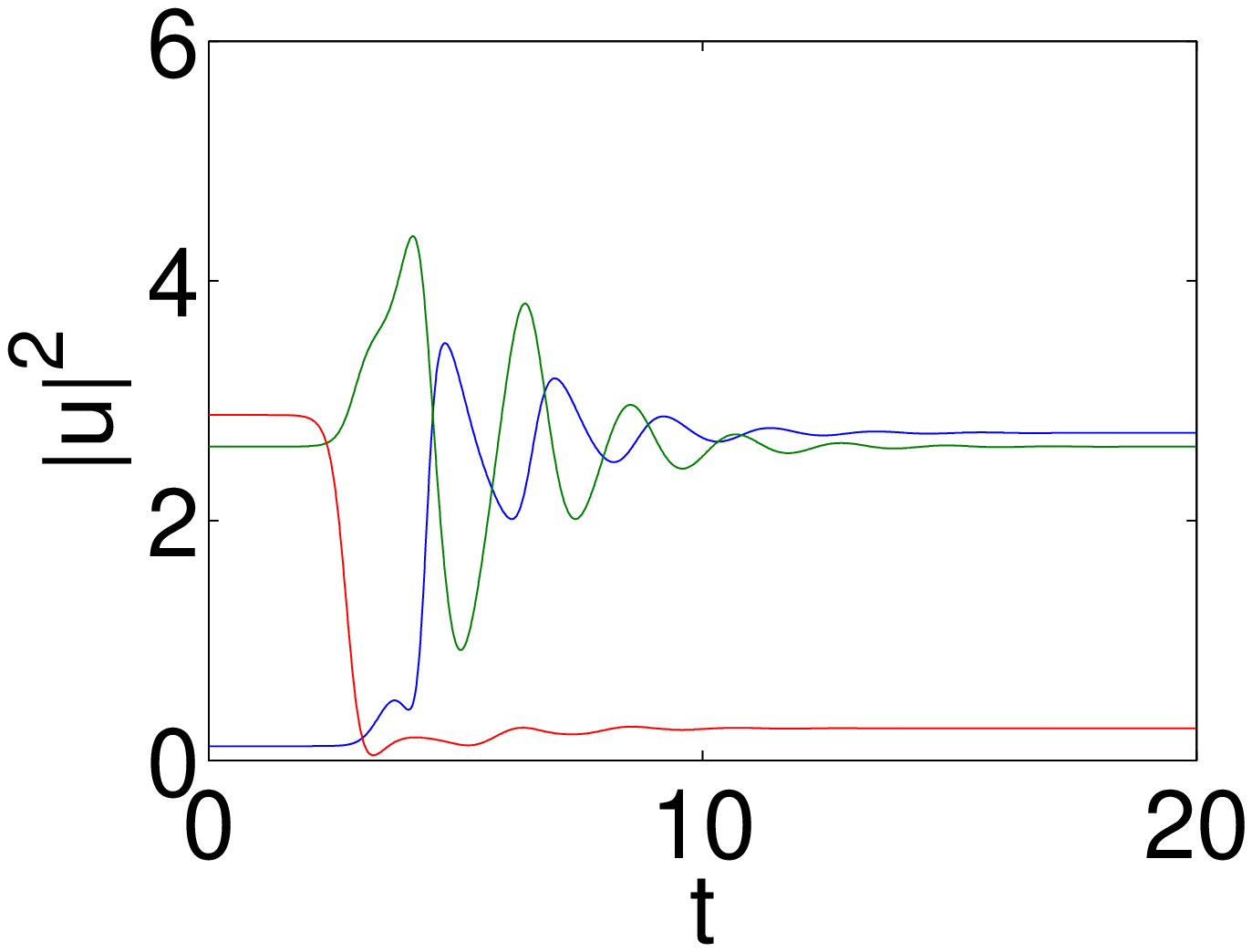}}}
    \caption{The time evolution plots of the trimer case II with $\epsilon=1$, $\rho_r=-1$ and $\rho_{im}=1$ when $\gamma=3$. The two unstable branches
(red diamonds and black squares) tend to a dynamically stable configuration
which is a mirror image of the red diamonds branch.}
    \label{fig12}
    \end{figure}

    \item Trimer Case III:

    Finally, we turn to a consideration, using the same parametric setting
as in case II, of the numerical results by solving Eq.~(\ref{trimercase3})
for case III in Figs.~\ref{fig13}-\ref{fig15}.
Four distinct branches of solutions
are observed in this case. The branches denoted by red diamonds and black
squares exist only for small values of
the linear gain/loss parameter $\gamma$,
are stable and terminate at $\gamma=0.65$. The other two branches, namely the blue stars and the green circles collide and terminate at $\gamma=2.1$. The
green circles branch is unstable in this case, due to a complex quartet of
eigenvalues (observed in Fig.~\ref{fig14}). On the other hand, the blue
stars' branch is stable up to $\gamma=2.05$ a critical point at which
branches of case II (the red diamonds branch referred to
in case II as having a pitchfork bifurcation at the same value and its
mirror symmetric image) emerge.
Notice that this detail is not discernible in the eigenvalue plots
of Fig.~\ref{fig13}.

    \begin{figure}[htp]
    \scalebox{0.5}{\includegraphics{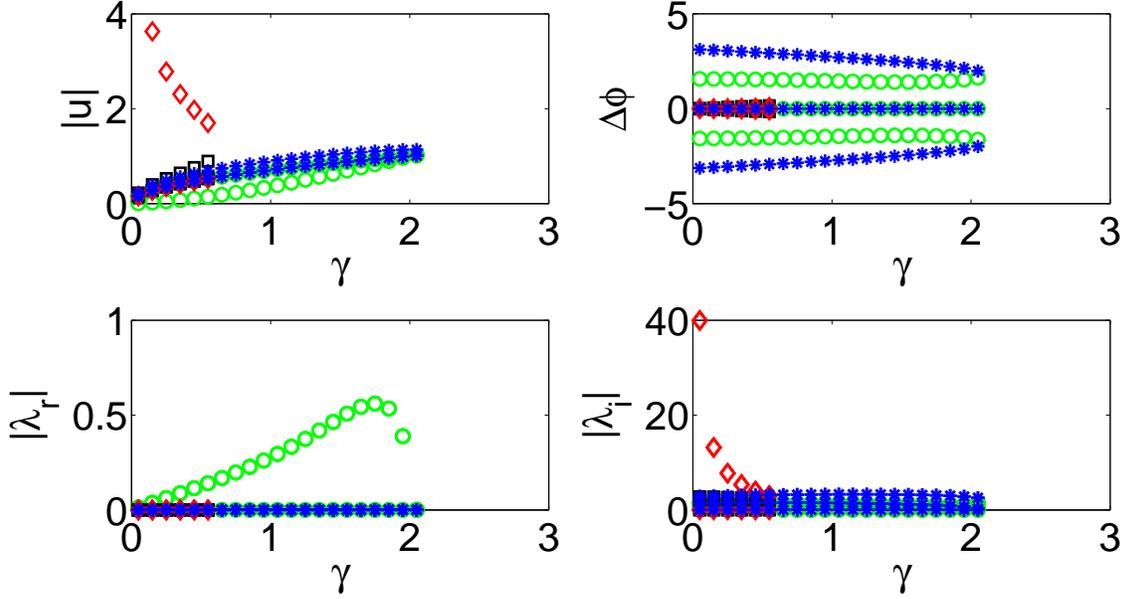}}
    \caption{The solution profile for the nonlinear-PT-symmetric
trimer of Case III (special symmetric solutions)
with $\epsilon=1$, $\rho_r=-1$ and $\rho_{im}=1$. The norms
are not squared in the top left panel to improve the visibility of the
branches (given the disparity of the relevant amplitudes).
    The four branches are denoted by blue stars, red diamonds,
black squares and green circles.
    The blue stars, red diamonds and black squares branches always have two pairs of purely imaginary eigenvalues, while
    the green circles branch always has a complex quartet. The blue stars
branch terminates with the green circles at $\gamma=2.1$,
    while the red diamonds and black squares terminate together
at $\gamma=0.65$.}
    \label{fig13}
    \end{figure}

    \begin{figure}[htp]
    \scalebox{0.4}{\includegraphics{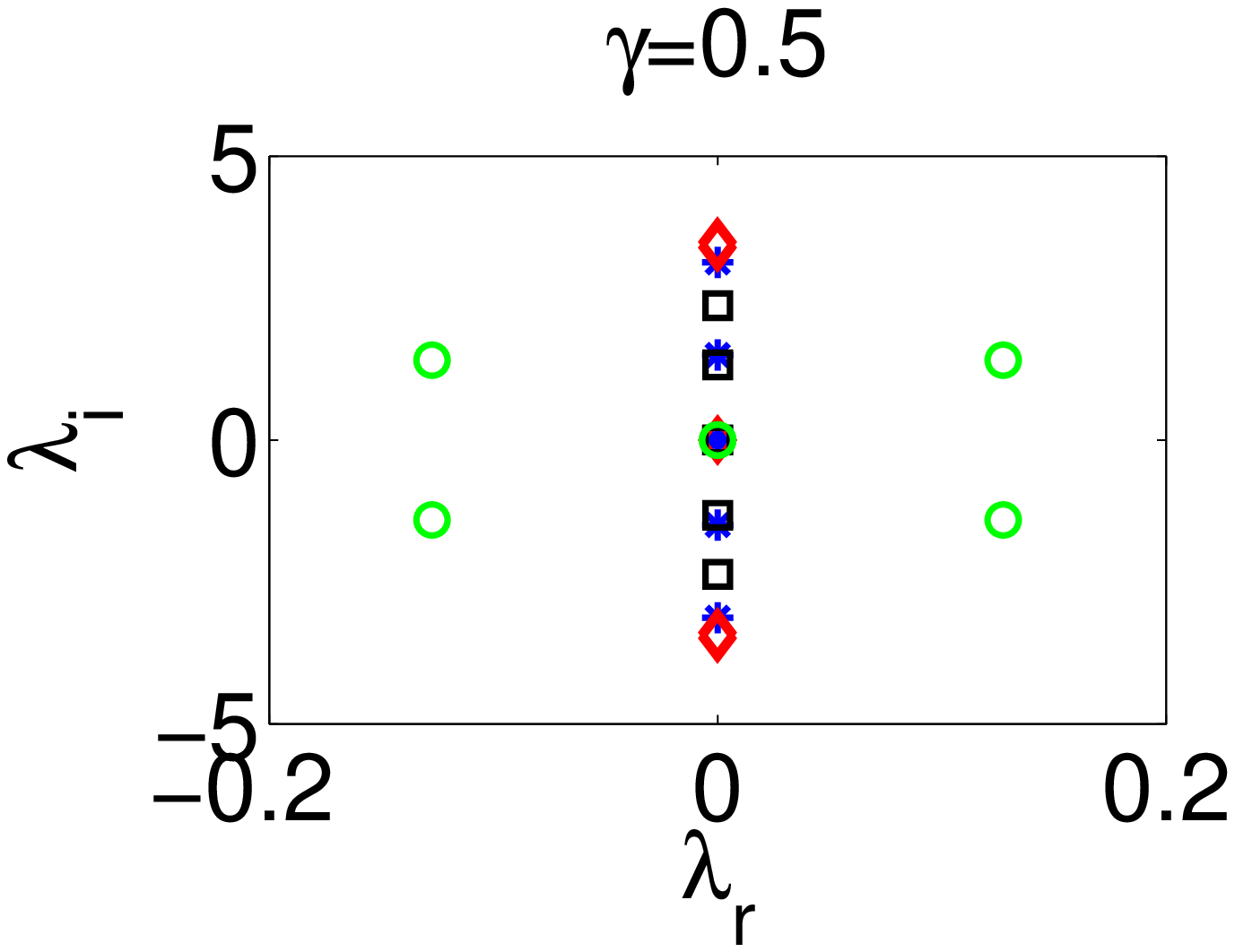}}
    \scalebox{0.4}{\includegraphics{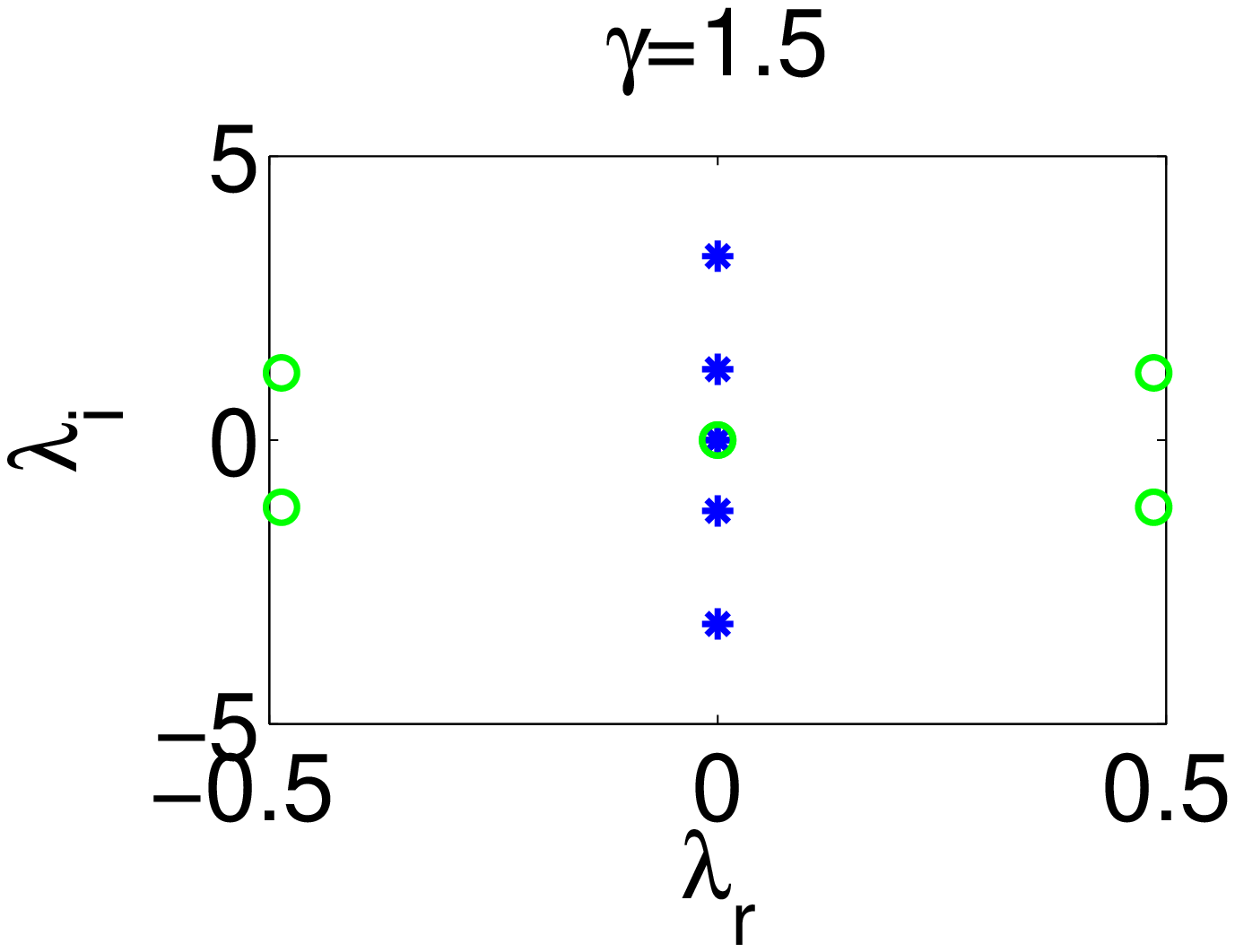}}
    \caption{The plots of the spectral plane of the linear stability
eigenvalues for the nonlinear-PT-symmetric trimer Case III
(special symmetric solutions)
with $\epsilon=1$, $\rho_r=-1$ and $\rho_{im}=1$.}
    \label{fig14}
    \end{figure}

The dynamics of the different configurations are shown in Fig.~\ref{fig15}.
The blue stars, red diamonds and black squares special symmetric branches
of solutions are stable and thus preserve their shape. On the other hand,
the green circles for $\gamma=0.5$ are subject to the oscillatory
instability predicted by the linear stability analysis. This, in turn,
results into long-lived oscillatory dynamics of the system, as indicated
in the bottom right of  Fig.~\ref{fig15}.

    \begin{figure}[htp]
    \subfigure[\ blue stars branch]{\scalebox{0.4}{\includegraphics{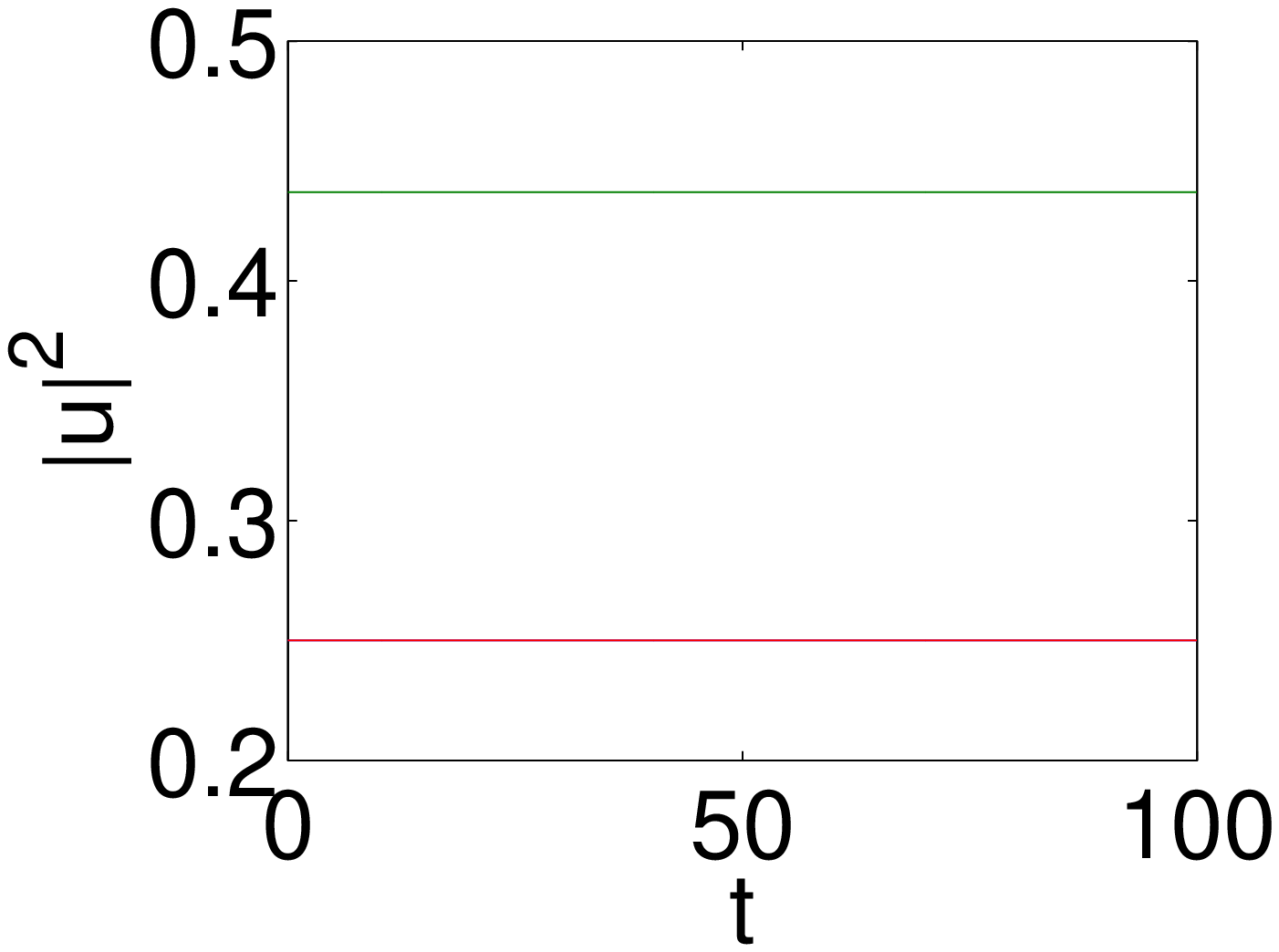}}}
    \subfigure[\ red diamonds branch]{\scalebox{0.4}{\includegraphics{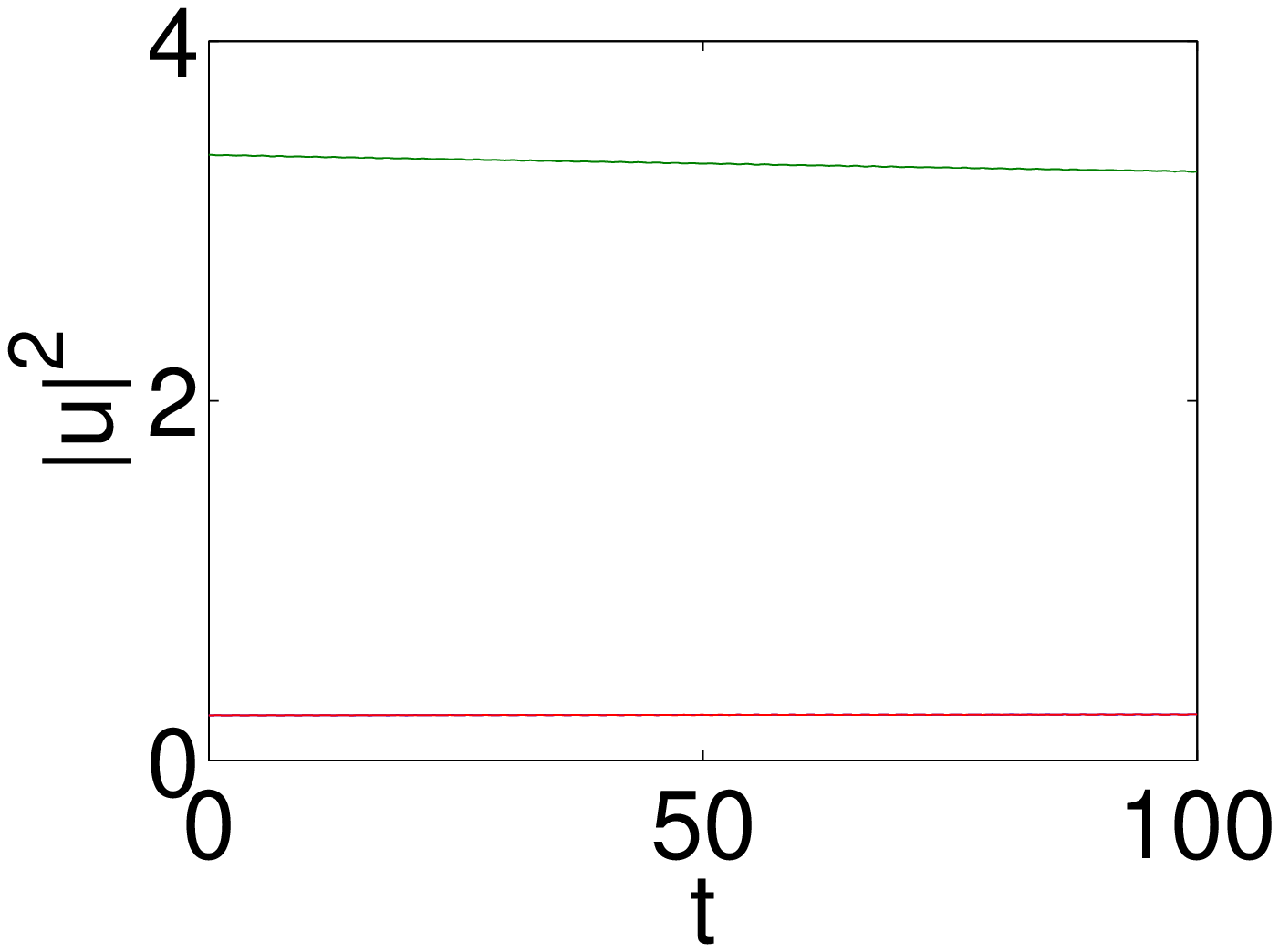}}}
    \subfigure[\ black squares branch]{\scalebox{0.4}{\includegraphics{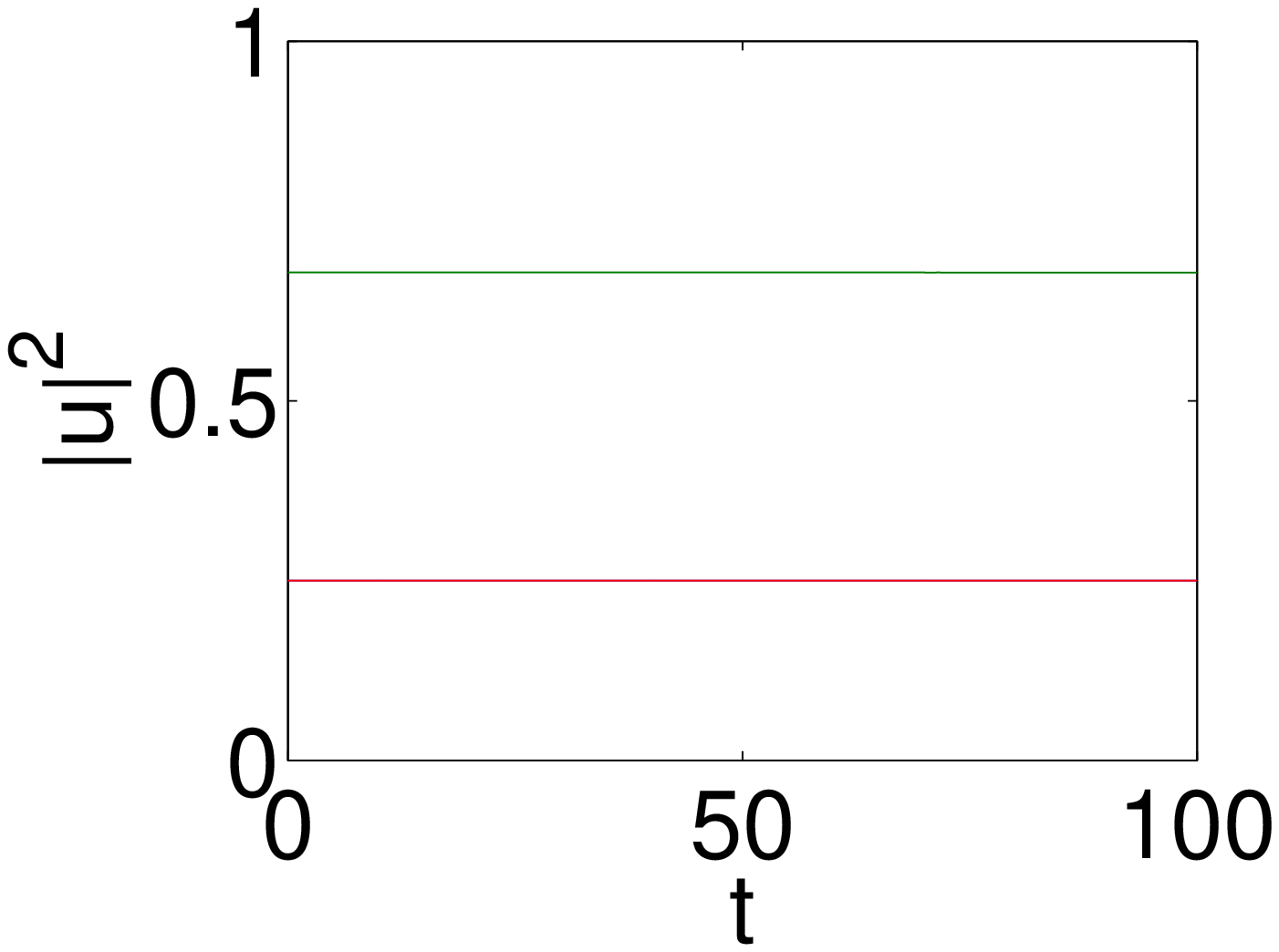}}}
    \subfigure[\ green circles branch]{\scalebox{0.4}{\includegraphics{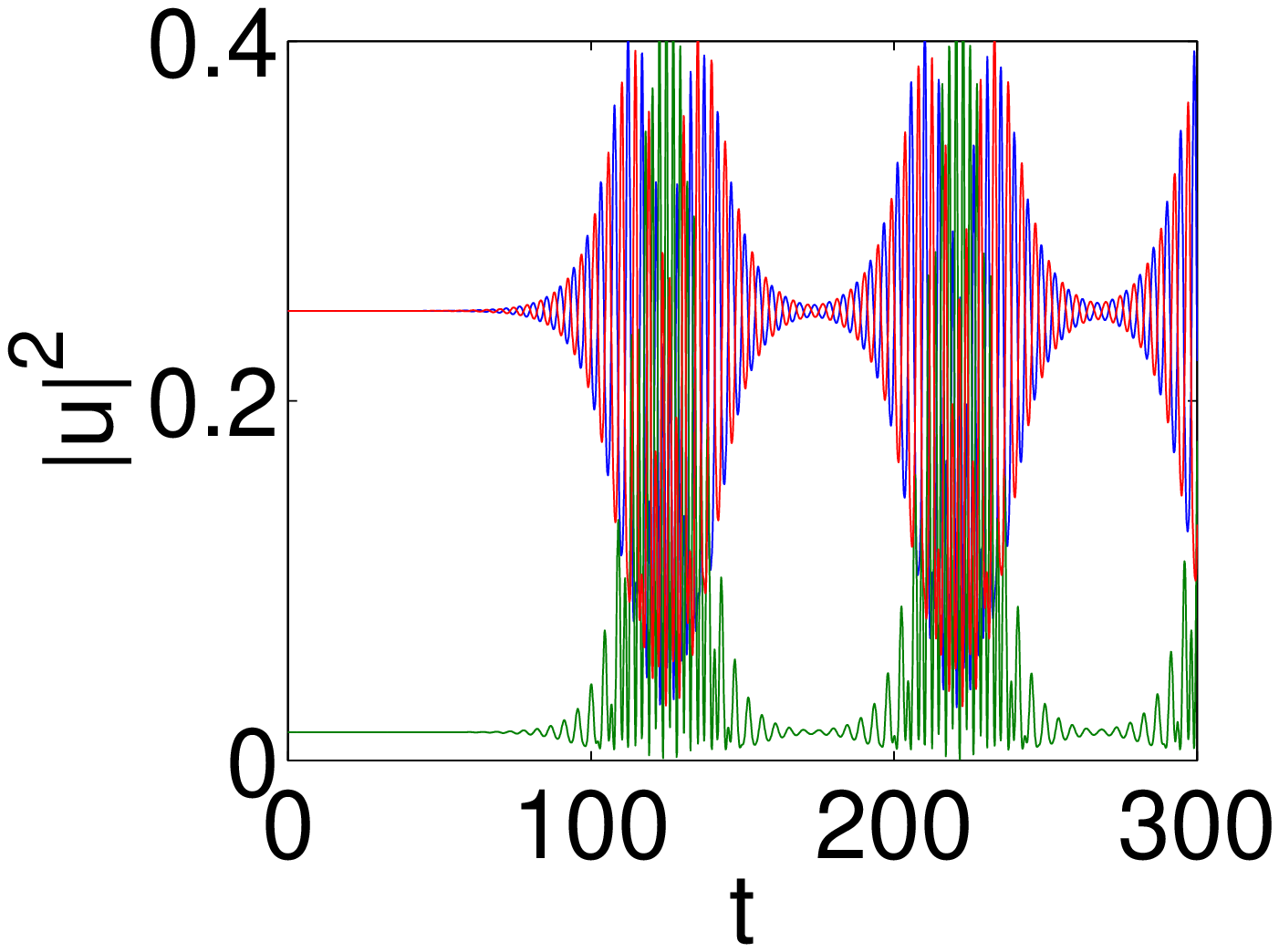}}}
    \caption{The time evolution plots of the nonlinear-PT-symmetric
trimer in Case III of special symmetric solutions
with $\epsilon=1$, $\rho_r=-1$ and $\rho_{im}=1$ when $\gamma=0.5$.
The only unstable configuration is the green circle one of the bottom right
which leads to long-lived oscillatory dynamics.}
    \label{fig15}
    \end{figure}
    \end{itemize}

\section{The Special Case of Linear-PT-Symmetric Oligomers}

As a special case with $\rho_r=-1$ and $\rho_{im}=0$ of what we did above,
the linear PT-symmetric oligomer case example
has been addressed in the earlier work
of~\cite{pgk}.


\begin{figure}[htp]
\scalebox{0.5}{\includegraphics{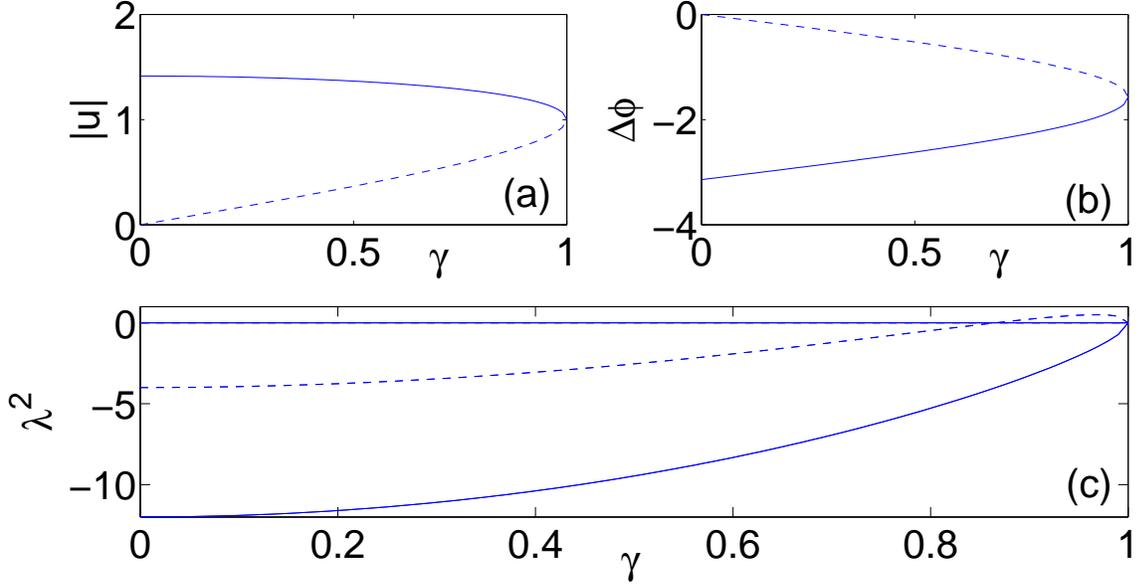}}
\caption{(Color online) The two branches of solutions for the dimer problem are shown
for parameter values $\epsilon=E=1$. (a) The amplitude
of the sites,(b) their relative phase, and (c)
the (nontrivial) squared eigenvalue of the two branches.
The solid line
corresponds to the always stable branch, while the
dashed line corresponds to the branch, which acquires
a real eigenvalue pair above a certain $\gamma=\sqrt{\epsilon^2-E^2/4}$.
Reprinted with permission from~\cite{pgk}.}
\label{dimer}
\end{figure}

Fig.~\ref{dimer} shows the profile of the two branches, which are analogues
of the nonlinear-PT-symmetric dimer  displayed in Fig.~\ref{fig4}.
The branch denoted by dashed line corresponds to the blue stars branch
of case I within
the nonlinear-PT-symmetric dimer with $(-)$ sign in
Eq.~(\ref{Asquared_A_equal_B_dimer_case}) and is stable when
$\gamma^2 \le k^2-E^2/4$, whereas the solid line branch corresponding to the
red diamonds branch (of case I in the nonlinear-PT-symmetric dimer)
is always stable.
The linearization around these branches can be performed explicitly in this
simpler linear-PT-symmetric case
yielding the nonzero eigenvalue pairs
$\pm 2 i \sqrt{2(\epsilon^2-\gamma^2)-E\sqrt{\epsilon^2-\gamma^2}}$
for the first and $\pm 2 i \sqrt{2(\epsilon^2-\gamma^2)+E\sqrt{\epsilon^2-\gamma^2}}$ for the
second (notice that the latter can never become real).

It is relevant to note here that the two branches ``die'' in a
saddle-center bifurcation at $\gamma=\epsilon$, as shown in the figure.
This point coincides with the linear PT-symmetric dimer
phase transition.
As indicated before, in the nonlinear dimer, the two branches die
when the restriction (\ref{dimer1_restriction}) is no longer satisfied.
Nevertheless, the nonlinear solutions of the latter case can generally
exist past the linear phase transition (and even arbitrarily past
that as in the case II solutions) and moreover asymmetric solutions can exist
due to the interplay of linear and nonlinear gain/loss a feature absent
in the simpler linear-PT-symmetric dimer.

\begin{figure}[htp]
\scalebox{0.5}{\includegraphics{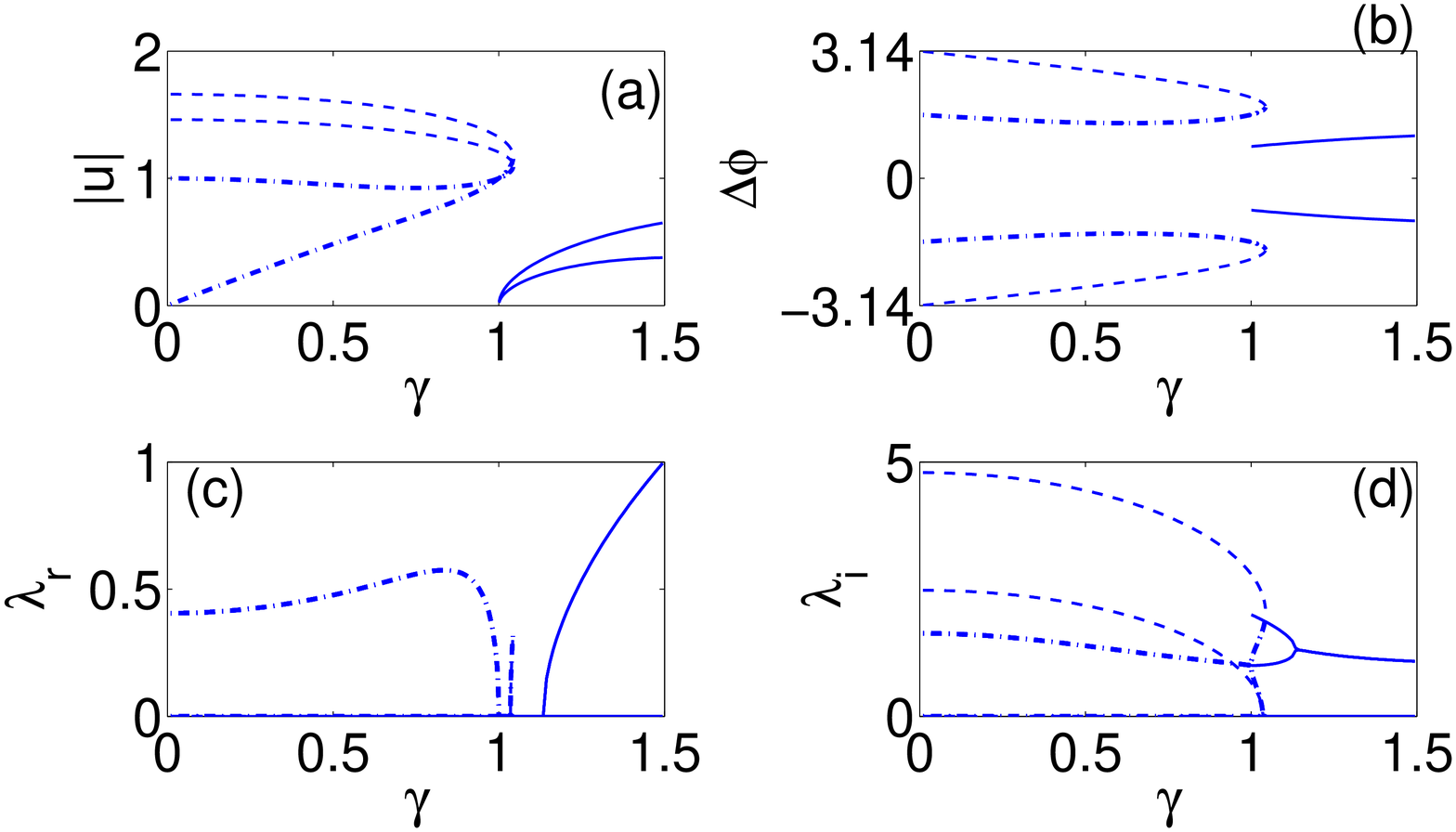}}
\caption{(Color online) Existence and stability of solutions for the case of the
trimer analogous to Fig.~\ref{dimer} with parameters $E=1$ and
$\phi_b$ normalized to $0$ (without loss of generality). There are three branches:
$u^{(1)}$ (dash-dotted line),
$u^{(2)}$ (dashed line), and $u^{(3)}$ (solid line).
For each branch, two curves in (a) stand for $A,B,C$ (since $C=A$), and two curves
in (b) stand for $\phi_a$ and $\phi_c$.
Panels (c) and (d) are the illustration of both real and imaginary
parts for the eigenvalues. There exist two solutions (the dashed and the
dash-dotted) one of which is stable (dashed) and one generically unstable
(dashed-dotted) which collide and disappear for $\gamma=1.043$. The third
solid branch emerges from the limit $\gamma=1$ and persists thereafter
(for all values of $\gamma$ considered), although it becomes unstable for
$\gamma>1.13$. Reprinted with permission from~\cite{pgk}.
}
\label{trimer}
\end{figure}

As an analogue of the case I solutions of the nonlinear-PT-symmetric
trimer, we present a prototypical example of the branches that may arise in
the case of the linear-PT-symmetric
trimer in Fig.~\ref{trimer} for $E=\epsilon=1$.
There are three distinct branches. Two of them collide
in a saddle-center bifurcation (for $\gamma=1.043$) and disappear thereafter.
The other one emerges and bifurcates from zero amplitude for
$\gamma > \sqrt{2\epsilon^2-E^2}$ and persists beyond the critical point of
the linear PT phase transition
$\gamma_{PT}=\sqrt{2}\epsilon$, (although it is unstable in that regime). Hence the feature of solutions persisting past the linear
PT phase transition exists even in the linear-PT-symmetric trimer,
yet other more complex features do not appear in this setting. A canonical
example thereof is the existence of asymmetric solutions (which, in turn,
possess asymmetric spectra). The latter trait is amply present
in the nonlinear-PT-symmetric trimer of the previous section.

It should be added that in addition to the oligomers of the dimer
and the trimer variety, recently there has also been considerable
interest towards the study of quadrimer settings. Such settings
were again initiated in the examination of~\cite{pgk}, where
the particular (and simpler) case of $(-i \gamma, -i \gamma, + i\gamma,
+i \gamma)$ i.e., a linear-PT-symmetric quadrimer with two lossy sites
on the one side and two gain nodes on the other side was considered.
The considerations of~\cite{pgk} were generalized in the very recent
study of~\cite{konorecent2}, which examined the configuration
$(-i \gamma_1, -i \gamma_2, + i\gamma_2,
+i \gamma_1)$ i.e., the most general bi-parametric gain/loss family
of quadrimers. In both cases, the possibility of solutions that
emerge purely in the nonlinear regime and of ones that continue past
the critical point of the underlying linear PT phase transition was
identified. Another reason why such quadrimers are of interest is
that they can be thought of as the prototypical building blocks
(if placed on a square ``plaquette'') of a two-dimensional
PT-symmetric lattice~\cite{uwe}. To the best of our knowledge,
nonlinear-PT-symmetric quadrimers, along the lines considered
herein have not been examined to date.

\section{Conclusions and Future Challenges}

In the above
study, we illustrated some interesting characteristics which emanate from the
interplay of nonlinearity with PT-symmetric linear Hamiltonians in the
case of oligomer configurations.
The basic underlying premise which has been explained in
the recent works of~\cite{konorecent3,uwe} and has been
explored in numerous others such as~\cite{pgk,miron,konorecent3,uwe}
is that the nonlinear and the PT-symmetric-linear part of the system
at hand no longer commute and hence give rise to novel phenomenology
that is not expected for linear PT symmetric systems.
A principal element of this phenomenology that arises even
for PT-symmetric oligomers (trimers, quadrimers, etc.)
with merely linear gain/loss is the fact that nonlinear solutions
may exist that do not have a corresponding linear limit and
which, in fact, defy the threshold for the linear PT phase transition
in that they exist for arbitrary gain/loss parameter values past
that critical point. The introduction of a nonlinear
gain/loss pattern considered herein for dimers and trimers
(and earlier in a more cursory way in~\cite{miron} for dimers)
presents additional possibilities stemming from the interplay
of linear and nonlinear gain/loss profiles. These include
among others the emergence of asymmetric solutions which not
only are involved in symmetry breaking (pitchfork) bifurcations
but also produce asymmetric linearization matrices with spectral
properties that reflect this asymmetry.

We believe that this direction of studies is particularly intriguing
for further progress, especially as the complexity of the problem
increases within the confines of a full one-dimensional chain,
but also even for
fundamental two-dimensional entities, such as the quadrimer based
plaquettes. These themes constitute pristine territory for further
exploration at the theoretical level. Naturally, on the other hand,
a potential generalization of the earlier experiments of~\cite{kip}
towards the inclusion of nonlinear amplification and amplitude-dependent
loss in balance with each other would be most worthwhile to consider
in order to take advantage of the
considerable additional wealth of phenomenology of the latter system.

\acknowledgments PGK gratefully acknowledges support from
the US National Science Foundation (grant DMS-0806762),
the Alexander von Humboldt Foundation and the Alexander S.
Onassis Public Benefit Foundation.



\begin{thebibliography}{99}
\bibitem{bend} C.M. Bender and S. Boettcher,
Phys. Rev. Lett. {\bf 80}, 5243 (1998); C.M. Bender, S. Boettcher
and P.N. Meisinger, J. Math. Phys. {\bf 40}, 2201 (1999).


\bibitem{christo1} Z.H. Musslimani, K.G. Makris, R. El-Ganainy
and D.N. Christodoulides, Phys. Rev. Lett. {\bf 100}, 030402 (2008);
K.G. Makris, R. El-Ganainy, D.N. Christodoulides and Z.H. Musslimani,
Phys. Rev. A {\bf 81}, 063807 (2010).

\bibitem{usrecent} V. Achilleos, P.G. Kevrekidis, D.J. Frantzeskakis,
R. Carretero-Gonz{\'a}lez, arXiv:1202.1310.


\bibitem{salamo} A. Guo, G. J. Salamo, D. Duchesne, R. Morandotti,
  M. Volatier-Ravat, V. Aimez,
G. A. Siviloglou and D. N. Christodoulides,
  Phys. Rev. Lett. {\bf 103}, 093902 (2009).


\bibitem{kip} C.E. R{\"u}ter, K.G. Makris, R. El-Ganainy,
D.N. Christodoulides, M. Segev, D. Kip,
Nature Phys. {\bf 6}, 192 (2010).

\bibitem{tsampikos_recent} J. Schindler,
A. Li, M.C. Zheng, F.M. Ellis and T. Kottos,
Phys. Rev. A {\bf 84}, 040101 (2011).

\bibitem{kot1} H. Ramezani, T. Kottos, R. El-Ganainy and D.N.
Christodoulides, Phys. Rev. A {\bf 82}, 043803 (2010).

\bibitem{sukh1} A.A. Sukhorukov, Z. Xu and Yu.S. Kivshar,
Phys. Rev. A {\bf 82}, 043818 (2010).

\bibitem{kot2} M.C. Zheng, D.N. Christodoulides, R. Fleischmann
and T. Kottos, Phys. Rev. A {\bf 82}, 010103(R) (2010).

\bibitem{grae1} E.M. Graefe, H.J. Korsch and A.E. Niederle,
Phys. Rev. Lett. {\bf 101}, 150408 (2008).

\bibitem{grae2} E.M. Graefe, H.J. Korsch and A.E. Niederle,
Phys. Rev. A {\bf 82}, 013629 (2010).

\bibitem{kot3} Z. Lin, H. Ramezani, T. Eichelkraut, T. Kottos,
H. Cao and D.N. Christodoulides, Phys. Rev. Lett. {\bf 106}, 213901 (2011).

\bibitem{pgk} K. Li and P. G. Kevrekidis
Phys. Rev. E {\bf 83}, 066608 (2011)

\bibitem{dmitriev1} S.V. Dmitriev, S.V. Suchkov, A.A. Sukhorukov,
and Yu.S. Kivshar,
Phys. Rev. A {\bf 84}, 013833 (2011)

\bibitem{dmitriev2} S.V. Suchkov,  B.A. Malomed, S.V. Dmitriev and
Yu.S. Kivshar, Phys. Rev. E 84, 046609 (2011).

\bibitem{R30add1} R. Driben and B. A. Malomed, Opt. Lett. {\bf 36}, 4323 (2011).

\bibitem{R30add2} R. Driben and B. A. Malomed, Europhys. Lett. {\bf 96}, 51001
(2011).

\bibitem{R30add3} F. Kh. Abdullaev, V.V. Konotop, M. \"Ogren and M. P.
S{\o}rensen, Opt. Lett. {\bf 36}, 4566 (2011).

\bibitem{R30add4} N.V. Alexeeva, I.V. Barashenkov, A.A. Sukhorukov, and 
Yu.S. Kivshar, Phys. Rev. A {\bf 85}, 063837 (2012).

\bibitem{R30add5} A.A. Sukhorukov, S.V. Dmitriev and Yu.S. Kivshar,
Opt. Lett. {\bf 37}, 2148 (2012).

\bibitem{R34} H. Cartarius and G. Wunner, Phys. Rev. A {\bf 86}, 013612 (2012);
see also arXiv:1207.1669, J. Phys. A, in press (2012).

\bibitem{R44} E.-M. Graefe, arXiv:1206.4806.


\bibitem{R46} A.S. Rodrigues, K. Li, V. Achilleos, P.G. Kevrekidis, D.J. Frantzeskakis, C.M. Bender,
arXiv:1207.1066.


\bibitem{miron} A.E. Miroshnichenko, B.A. Malomed, and Yu.S. Kivshar
Phys. Rev. A {\bf 84}, 012123 (2011).

\bibitem{konorecent} F.Kh. Abdullaev, Y.V. Kartashov, V.V. Konotop
and D.A. Zezyulin, Phys. Rev. A {\bf 83}, 041805 (2011)


\bibitem{konorecent2} D. A. Zezyulin, Y. V. Kartashov, V. V. Konotop,
Europhys. Lett. {\bf 96}, 64003 (2011).

\bibitem{konorecent3} D. A. Zezyulin, V. V. Konotop,
Phys. Rev. Lett. {\bf 108}, 213906 (2012).

\bibitem{majoh} M. Johansson,
J. Phys. A: Math. Gen.  {\bf 37}, 2201 (2004);
R.H. Goodman, J. Phys. A: Math. Theor. {\bf 44}, 425101 (2011).

\bibitem{todd} T. Kapitula, P. Kevrekidis, and Z. Chen,
SIAM J. Appl. Dyn. Sys. {\bf 5}, 598 (2006).

\bibitem{uwe} K. Li, P.G. Kevrekidis, B.A. Malomed and U. G{\"u}nther,
Nonlinear PT-symmetric plaquettes, 
arXiv:1204.5530, J. Phys. A, in press (2012).

\end{thebibliography}
\end{document}